%% file: paperv2.tex
\documentclass[aps,
prd,
twocolumn,
superscriptaddress,
lengthcheck,showpacs,letterpaper,nofootinbib,
floatfix]{revtex4-1}
\usepackage[pdftex]{graphicx}
\usepackage{natbib}
\usepackage{lipsum}
\usepackage{placeins}
\usepackage{dcolumn}
\usepackage{multirow}
\usepackage{amsmath}
\usepackage{amsthm, amsfonts, amssymb}
\usepackage{array}
\usepackage{times}
\usepackage{latexsym}
\usepackage{hyperref}
\hypersetup{backref,  
colorlinks=true,
linkcolor=red,
filecolor=red,
citecolor=blue}
\usepackage{epsfig}
\usepackage{subfigure}
\usepackage{bm}
\usepackage[usenames,dvipsnames]{color}
\usepackage[normalem]{ulem}

\newcommand{\ii}{{\rm i}}
\newcommand{\D}{\mathrm{d}}
\newcommand{\eff}{\mathrm{eff}}

\newcommand{\real}{\mathrm{real}}

\newcommand{\EOB}{\mathrm{EOBNRv2}}

\newcommand{\M}{\mathit{M}}

\newcommand{\bnk}{\mathrm{bank}}

\newcommand{\mn}{\mathrm{min}}
\newcommand{\mx}{\mathrm{max}}
\newcommand{\tr}{\mathrm{tr}}

\newcommand{\Hyb}{\mathrm{Hyb}}
\newcommand{\leftn}{\left|\left|}
\newcommand{\rightn}{\right|\right|}

\newcommand{\Mis}{\mathcal{M}}
\newcommand{\Olap}{\mathcal{O}}
\newcommand{\FF}{\mathrm{FF}}

\newcommand{\N}{\mathrm{N}}
\newcommand{\cyc}{\mathrm{cyc}}

\makeatletter
\newcommand{\etal}{\textit{et~al}\@ifnextchar{\relax}{.\relax}{\ifx\@let@token.\else\ifx\@let@token~.\else.\@\xspace\fi\fi}}
\makeatother

\def\l({\left(}
\def\r){\right)}

\newcommand{\CITA}{\affiliation{Canadian Institute for Theoretical
    Astrophysics,
 University of Toronto,
    Toronto, ON M5S 3H8, Canada}} %
\newcommand{\CIFAR}{\affiliation{Canadian Institute for Advanced Research, 180 Dundas St.~West, Toronto, ON M5G 1Z8, Canada}} %
\newcommand{\DAA}{\affiliation{Department of Astronomy and Astrophysics, 
University of Toronto, Toronto, ON M5S 3H4, Canada}}
\newcommand{\SYR}{\affiliation{Department of Physics, Syracuse University, Syracuse, NY 13244, USA}}
\newcommand{\CIT}{\affiliation{Theoretical Astrophysics 350-17, California Institute of Technology, Pasadena, California 91125, USA}}
\newcommand{\LIGOCIT}{\affiliation{LIGO Laboratory, California Institute of Technology, Pasadena CA 91125, USA}}

\newcommand{\Cornell}{\affiliation{Center for Radiophysics and Space
    Research, Cornell University, Ithaca, New York 14853, USA}}

\begin{document}

\author{Prayush Kumar}\thanks{prayush.kumar@ligo.org}
\SYR
\author{Ilana MacDonald}
\CITA\DAA
\author{Duncan A. Brown}
\SYR\LIGOCIT
\author{Harald P. Pfeiffer}
\CITA\CIFAR
\author{Kipp Cannon}
\CITA

\author{Michael Boyle} \Cornell 
\author{Lawrence E.~Kidder} \Cornell 
\author{Abdul H.~Mrou\'{e}} \CITA 
\author{Mark A.~Scheel} \CIT 
\author{B\'{e}la Szil\'{a}gyi} \CIT 
\author{An\i l Zengino\u{g}lu} \CIT 

\date{\today}
\title{Template Banks for Binary black hole searches with Numerical Relativity waveforms}
\begin{abstract}
      Gravitational waves (GW) from coalescing stellar-mass black hole
	  binaries (BBH) are expected to be detected by the Advanced Laser
	  Interferometer Gravitational-wave Observatory and Advanced Virgo.
	  Detection searches operate by matched-filtering the detector data 
	  using a bank of waveform templates.
	  Traditionally, template banks for BBH are constructed from
	  intermediary analytical waveform models which are calibrated
	  against numerical relativity simulations and which can be
	  evaluated for any choice of BBH parameters.  This paper explores
	  an alternative to the traditional approach, namely the
	  construction of template banks directly from numerical BBH
	  simulations.  Using non-spinning BBH systems as an example, we
	  demonstrate which regions of the mass-parameter plane can be
	  covered with existing numerical BBH waveforms.  We estimate the
	  required number and required length of BBH simulations to cover
	  the entire non-spinning BBH parameter plane up to mass-ratio 10,
	  thus illustrating that our approach can be used to guide parameter placement of future numerical simulations.   We
	  derive error bounds which are independent of analytical waveform
	  models; therefore, our formalism can be used to independently test
	  the accuracy of such waveform models. The resulting template banks 
	  are suitable for advanced LIGO searches.

\end{abstract}
\maketitle
\section{Introduction}

\input{introduction.tex}

\section{Waveforms}\label{s1:waveforms}

In the sections that follow, we will describe the construction of template 
banks for NR or NR-PN hybrid waveform templates. The NR waveforms that we
use correspond to mass ratios $q=\{1,2,3,4,6,8\}$, and were simulated using
the SpEC code~\cite{spec}. The construction of hybrid
waveforms involves joining a long inspiral portion, modeled using PN theory,
to the merger-ringdown waveform from NR. In this section we describe both, the 
NR waveforms and the PN models used in our study. Measuring the effectualness
of these banks involves simulating a population of BBH signals. We use the 
recently published EOBNRv2 model~\cite{BuonannoEOBv2Main} to obtain waveforms
for BBHs with arbitrary masses. This model was calibrated against five out of 
the six NR simulations we use to construct our banks, and is expected to be 
faithful at comparable mass ratios~\cite{BuonannoEOBv2Main}. In this section, 
we briefly summarize the construction of EOBNRv2 waveforms as well.

\subsection{Numerical Relativity simulations}\label{s2:NRwaveforms}
\input{nrwaveformsdescription.tex}

\subsection{Post-Newtonian waveforms}\label{s2:PNwaveforms}

\input{pndescription.tex}

\subsection{PN-NR hybrid waveforms}\label{s2:NRpNhybridwaveforms}
\input{hybridizationprocedures.tex}

\subsection{Effective-One-Body model}\label{s2:EOBwaveforms}

\input{eobnrv2description.tex}

\section{Quantifying waveform accuracy \& bank effectualness}\label{s1:quantifyingerrors}
\input{mathdefinitions.tex}

\section{Constructing a template bank for NR waveforms}\label{s1:NRonlybank}
\input{nrbankresults.tex}

\section{Constructing a template bank for NR-PN hybrids}\label{s1:NRpNhybridbank}
\input{hybridbankresults.tex}

\section{Complete NR-PN hybrid bank for non-spinning BBH}\label{s1:futureNRpNhybridbank}
\input{futurehybridbankresults.tex}

\section{Conclusions}\label{s1:conclusions}
\input{conclusions.tex}

\acknowledgments


We thank Steve Privitera for useful code contributions and Ian Harry,
Alex Nitz, Stefan Ballmer and the Gravitational-Wave group at Syracuse
University for productive discussions. We also thank Thomas Dent for
carefully reading through the manuscript and providing feedback.
DAB, PK and HPP are grateful for hospitality of the TAPIR
group at the California Institute of Technology, where part of this
work was completed.  DAB and PK also thank the LIGO Laboratory
Visitors Program, supported by NSF cooperative agreement PHY-0757058,
for hospitality during the completion of this
work.
KC, IM, AHM and HPP acknowledge support by NSERC of Canada, the Canada
Chairs Program, and the Canadian Institute for Advanced Research.  We
further acknowledge support from National Science Foundation awards
PHY-0847611 (DAB and PK); PHY-0969111 and PHY-1005426 (MB, LEK); and
PHY-1068881, PHY-1005655, and DMS-1065438 (MAS, BS, AZ).  We are
grateful for additional support through a Cottrell Scholar award from the
Research Corporation for Science Advancement (DAB) and from the Sherman
Fairchild Foundation (MB, LEK, MAS, BS, AZ).
Simulations used in this work were performed with the \texttt{SpEC}
code~\cite{spec}.  Calculations were performed on the Zwicky cluster
at Caltech, which is supported by the Sherman Fairchild Foundation and
by NSF award PHY-0960291; on the NSF XSEDE network under grant
TG-PHY990007N; on the Syracuse University Gravitation and Relativity
cluster, which is supported by NSF awards PHY-1040231 and PHY-1104371
and Syracuse University ITS; and on the GPC supercomputer at the
SciNet HPC Consortium~\cite{scinet}. SciNet is funded by: the Canada
Foundation for Innovation under the auspices of Compute Canada; the
Government of Ontario; Ontario Research Fund--Research Excellence; and
the University of Toronto.  
\FloatBarrier
\bibliographystyle{apsrev4-1}
\bibliography{paper}

\end{document}

%% file: introduction.tex
Upgrades to the LIGO and Virgo observatories are
underway~\cite{Harry:2010zz,aVIRGO}, with first observation runs planned for
$2015$~\cite{Aasi:2013wya}. The construction of the Japanese detector KAGRA 
has also begun~\cite{Somiya:2011np}. The advanced detectors will be
sensitive to gravitational waves at frequencies down to 
$\sim 10$Hz, with an order of magnitude increase in sensitivity across the
band. This is a significant improvement over the lower cutoff of $40$Hz
for initial LIGO. Estimates for the expected rate of detection have
been placed between $0.4 - 1000$ stellar-mass binary black hole (BBH)
mergers a year~\cite{LSCCBCRates2010}. 
The uncertainty in these estimates comes from the uncertainties in the various
factors that govern the physical processes in the BBH formation 
channels~\cite{1973NInfo..27...86T,1973NInfo..27...70T}. 
In sub-solar metallicity environments, stars (in binaries) are expected to 
lose relatively less mass to stellar winds and form more massive remnants 
\cite{Webbink:1984ti,Kowalska:2012bb,Fryer:2011cx}. 
Population synthesis studies estimate that sub-solar metallicity environments
within the horizon of advanced detectors could increase the detection rates 
to be as high
as a few thousand per year~\cite{Dominik:2012kk,Belczynski:2012cx}. 
On the other hand, high recoil momenta during core-collapse and 
merger during the common-envelope phase of the binary star evolution
could also decrease the detection 
rates drastically~\cite{Fryer:2011cx,Dominik:2012kk}. 


Past GW searches  have focused on GW bursts~\cite{Abadie:2010mt,
Abadie:2010wx,Abadie:2012rq}; coalescing compact
binaries~\cite{Colaboration:2011nz,Abadie:2010yb,Abbott:2009qj,
Abbott:2009tt,Messaritaki:2005wv,Abadie:2011kd,Aasi:2012rja},
and ringdowns of perturbed black holes~\cite{Abbott:2009km}, amongst
others~\cite{Abbott:2003yq,Abbott:2005pu,Sintes:2005fp,Abadie:2011md,
Palomba:2012wn}. For coalescing BBHs, detection searches involve 
matched-filtering~\cite{Wainstein:1962,Allen:2005fk} of the instrument
data using large banks of theoretically modeled waveform templates
as filters~\cite{Sathyaprakash:1991mt,SathyaMetric2PN,OwenTemplateSpacing,
BabaketalBankPlacement,SathyaBankPlacementTauN,Cokelaer:2007kx}.
The matched-filter is the optimal linear filter to maximize the
signal-to-noise ratio (SNR), in the presence of stochastic 
noise~\cite{1057571}. It requires an accurate modeling of the gravitational 
waveform emitted by the source binary. Early LIGO-Virgo searches 
employed template banks of Post-Newtonian (PN) inspiral 
waveforms~\cite{Colaboration:2011nz,Abadie:2010yb,Abbott:2009qj,
Abbott:2009tt,Messaritaki:2005wv}, while more recent
searches targeting high mass BBHs used complete inspiral-merger-ringdown
(IMR) waveform templates~\cite{Abadie:2011kd,Aasi:2012rja}. 


Recent developments in Numerical Relativity (NR) have provided complete 
simulations of BBH dynamics in the strong-field regime, i.e. during the
late-inspiral and merger phases~\cite{Pretorius2005,Baker:2005vv,Campanelli:2005dd,
Pretorius2006,Lindblom:2005qh}. These simulations have contributed 
unprecedented physical insights to the understanding of BBH mergers
(see, e.g., ~\cite{Pretorius:2007nq,Hannam:2009rd,Hinder:2010vn,Pfeiffer:2012pc} 
for recent overviews of the field).
Due to their high computational cost, fully numerical simulations currently 
span a few tens of inspiral orbits before merger. For mass-ratios
$q=m_{1}/m_{2}=1,2,3,4,6,8$, the multi-domain Spectral Einstein code 
(SpEC)~\cite{spec} has been used to simulate 15--33 inspiral merger 
orbits~\cite{Buchman:2012dw,Mroue:2012kv,Mroue:2013inPrep}.
These simulations have been used to calibrate waveform models, for example,
within the effective-one-body (EOB) formalism~\cite{EOBOriginalBuonannoDamour,
EOBNRdevel01,BuonannoEOBv2Main,Taracchini:2012ig}. 
Alternately, inspiral waveforms from PN theory can be
joined to numerical BBH inspiral and merger waveforms, to construct longer 
\textit{hybrid} waveforms~\cite{Boyle:2011dy,MacDonald:2011ne,MacDonald:2012mp,
Ohme:2011zm,Hannam:2010ky}. NR-PN hybrids have been used to calibrate 
phenomenological waveform models~\cite{Ajith:2007qp,Santamaria:2010yb},
and within the NINJA project~\cite{Aylott:2009tn,Ajith:2012tt}
to study the efficacy of various GW search algorithms towards realistic (NR)
signals~\cite{Santamaria:2009tm,Aylott:2009ya}.

Constructing template banks for gravitational wave searches has
been a long sought goal for NR. Traditionally, intermediary waveform
models are calibrated against numerical simulations and then
used in template banks for LIGO searches~\cite{Abadie:2011kd,
Aasi:2012rja}. In this paper we explore an alternative to this
traditional approach, proposing the use of NR waveforms themselves
and hybrids constructed out of them as search templates.
For a proof of principle, we focus on the non-spinning BBH space, 
with the aim of extending to spinning binaries in future work. We
investigate exactly where in the mass space can the existing NR 
waveforms/hybrids be used as templates, finding that only six 
simulations are sufficient to cover binaries with 
$m_{1,2}\gtrsim 12M_\odot$ upto mass-ratio $10$. This method can
also be used as a guide for the placement of parameters for future
NR simulations. Recent work has shown that existing PN waveforms 
are sufficient for aLIGO searches for 
$M=m_1+m_2\lesssim 12M_\odot$~\cite{CompTemplates2009,Brown:2012nn}.
To extend the NR/hybrid bank coverage down to 
$M\simeq 12M_\odot$, we demonstrate that a total of $26$ 
simulations would be sufficient. The template banks are 
constructed with the requirement that the net SNR recovered for 
any BBH signal should remain above $96.5\%$ of its optimal value.
Enforcing this tells that that these $26$ simulations would be 
required to be $\sim 50$ orbits long. This goal is achievable,
given the recent progress in simulation 
technology~\cite{MacDonald:2012mp,Mroue:2013xna,BelaLongSimulation}. 
Our template banks are viable for GW searches with aLIGO, and the 
framework for using hybrids within the LIGO-Virgo software 
framework has been demonstrated in the NINJA-2
collaboration~\cite{NINJA2:2013inPrep}. In this paper, we also 
derive waveform modeling error bounds which are independent of
analytical models. These can be extended straightforwardly to 
assess the accuracy of such models.

First, we construct a bank for purely-NR templates, restricting to
currently available simulations~\cite{MacDonald:2012mp,Mroue:2012kv,
Buchman:2012dw,Mroue:2013xna,Mroue:2012kv}. We use a stochastic algorithm 
similar to Ref.~\cite{Harry:2009ea,Ajith:2012mn,Manca:2009xw}, 
and place a template bank grid
constrained to $q=m_1/m_2=\{1,2,3,4,6,8\}$. The bank placement 
algorithm uses the EOB model from Ref.~\cite{BuonannoEOBv2Main} (EOBNRv2), 
which was calibrated against NR for 5 out of these 6 mass-ratios. 
To demonstrate the efficacy of the 
bank, we measure its fitting-factors (FFs)~\cite{FittingFactorApostolatos} over
the BBH mass space. We simulate a population of $100,000$ BBH waveforms with
masses sampled uniformly over 
$3M_\odot\leq m_{1,2}\leq 200M_\odot$ and $M=m_1+m_2\leq 200M_\odot$, and filter
them through the template bank to characterize its SNR recovery. For a
bank of NR templates, any SNR loss accrued will be due to the coarseness
of the bank grid. We measure this requiring both signals and templates
to be in the same manifold, using the EOBNRv2 model for both. We find 
that for systems with chirp mass 
$\mathcal{M}_c \equiv (m_1 + m_2)^{-1/5} (m_1 m_2)^{3/5}$ above 
$\sim 27M_{\odot}$ and $1\leq q\leq 10$, this bank has FFs $\geq 97\%$ and
is sufficiently accurate to be used in GW searches.
We also show that the coverage of the purely-NR bank can be extended to
include $10\leq q\leq 11$, if we instead constrain it to templates with
mass-ratios $q=\{1,2,3,4,6,9.2\}$.

Second, we demonstrate that currently available PN-NR hybrid waveforms can be 
used as templates to search for BBHs with much lower masses. The hybrids
used correspond to mass-ratios $q=\{1,2,3,4,6,8\}$. We use two distinct methods
of bank placement to construct a bank with these mass-ratios, and compare the
two. The first method is the stochastic algorithm we use for purely-NR 
templates. The second is a deterministic algorithm, that constructs the 
two-dimensional bank (in $M$ and $q$) through a union of six one-dimensional
banks, placed separately for each allowed value of mass-ratio. Templates are
placed over the total mass dimension by requiring that all pairs of neighboring
templates have the same noise weighted overlap. As before, we measure the SNR 
loss from both banks, due to the discrete placement of the templates, by 
simulating a population of $100,000$ BBH signals, to find the SNR recovered.
We measure the intrinsic hybrid errors using the method of
Ref.~\cite{MacDonald:2011ne,MacDonald:2012mp}, and subsequently account for 
them in the SNR recovery fraction. We find that the NR-PN
hybrid bank is effectual for detecting BBHs with $m_{1,2}\geq 12M_{\odot}$, 
with FFs $\geq 96.5\%$. The number of templates required was found
to be close to that of a bank constructed using the second-order TaylorF2
hexagonal bank placement algorithm~\cite{Sathyaprakash:1991mt,SathyaMetric2PN,
OwenTemplateSpacing,BabaketalBankPlacement,
SathyaBankPlacementTauN,Cokelaer:2007kx}. We note that by pre-generating the
template for the least massive binary for each of the mass-ratios that 
contribute to the bank, we can re-scale it on-the-fly to different total 
masses in the frequency domain~\cite{Sathyaprakash:2000qx}. 
Used in detection searches, such a bank would be computationally inexpensive 
to generate relative to a bank of time-domain modeled waveforms.

Finally, we determine the minimal set of NR simulations that we would need to 
extend the bank down to $M\simeq 12M_\odot$. We find that a bank that
samples from the set of $26$ mass-ratios listed in Table~\ref{table:fullqlist}
would be sufficiently dense, even at the lowest masses, for binaries with 
mass-ratios $1\leq q\leq 10$. We show that this bank recovers more than
$98\%$ of the optimal SNR, not accounting for hybrid errors. 
To restrict the loss in event detection 
rate below $10\%$, we restrict the total SNR loss below $3.5\%$. 
This implies the hybrid error mismatches stay below $1.5\%$, which 
constrains the length of the NR part for each hybrid.
We find that NR simulations spanning about $50$ orbits of late-inspiral, merger
and ringdown would suffice to reduce the PN truncation error to the desired 
level. With such a bank of NR-PN hybrids and purely-PN templates for lower
masses, we can construct GW searches for stellar-mass BBHs with mass-ratios 
$q\leq 10$.

The paper is organized as follows, in Sec.~\ref{s2:NRwaveforms}, we discuss
the NR waveforms used in this study, in Sec.~\ref{s2:PNwaveforms} we
describe the PN models used to construct the NR-PN hybrids,  and in Sec.~\ref{s2:NRpNhybridwaveforms} we describe the
construction of hybrid waveforms. In Sec.~\ref{s2:EOBwaveforms} we 
describe the EOB model that we use to place and test the template
banks. In Sec.~\ref{s1:quantifyingerrors} we describe the accuracy
measures used in quantifying the loss in signal-to-noise ratio in a
matched-filtering search when using a discrete bank of templates and
in the construction of hybrid waveforms. In Sec.~\ref{s1:NRonlybank}
we describe the construction and efficacy of the NR-only banks, while in
Sec.~\ref{s1:NRpNhybridbank} we discuss the same for the NR-PN hybrid 
template banks constructed with currently available NR waveforms. In
Sec.~\ref{s1:futureNRpNhybridbank}, we investigate the parameter and length
requirements for future NR simulations in order to cover the entire 
non-spinning parameter space with $12 M_\odot\leq M\leq 200M_\odot$, 
$m_{1,2} \geq 3M_\odot$, and $1 \leq q \leq 10$. Finally, in 
Sec.~\ref{s1:conclusions} we summarize the results.

%% file: nrwaveformsdescription.tex
The numerical relativity waveforms used in this paper were produced
with the SpEC code~\cite{spec}, a multi-domain pseudospectral code to solve
Einstein’s equations. SpEC uses Generalized Harmonic coordinates,
spectral methods, and a flexible domain decomposition, all of which
contribute to it being one of the most accurate and efficient codes
for computing the gravitational waves from binary black hole
systems. High accuracy numerical simulations of the late-inspiral,
merger and ringdown  of coalescing binary black-holes have been
recently performed for mass-ratios $q\equiv m_1/m_2\in\{1,2,3,4,6,8\}$
~\cite{Buchman:2012dw,Scheel:2008rj,NRPNComparisonBoyleetal,Mroue:2012kv}.

The equal-mass,
non-spinning waveform covers 33 inspiral orbits and was first discussed
in~\cite{MacDonald:2012mp,Mroue:2012kv}. 
This waveform was obtained with numerical techniques similar to those 
of~\cite{Buchman:2012dw}. The unequal-mass waveforms of mass ratios 
$2, 3, 4,$ and $6$ were presented in detail in Ref.~\cite{Buchman:2012dw}.
The simulation with mass ratio $6$ covers about 20 orbits and the
simulations with mass ratios 2, 3, and 4 are somewhat shorter and
cover about 15 orbits. The unequal mass waveform with mass ratio 8 was
presented as part of the large waveform catalog
in~\cite{Mroue:2013xna,Mroue:2012kv}. It is approximately 25 orbits in
length. 
We summarize the NR simulations used in this study in 
Table~\ref{table:etalist4}.

\begin{table}
\begin{tabular}{| c | c | c |}
\hline
$\eta$ & q & Length (in orbits)\\ \hline
0.25 & 1 & 33 \\
0.2222 & 2 & 15 \\
0.1875 & 3 & 18 \\
0.1600 & 4 & 15 \\
0.1224 & 6 & 20 \\
0.0988 & 8 & 25 \\
\hline
\end{tabular}
\caption{SpEC BBH simulations used in this study.  Given are symmetric mass-ratio $\eta$, mass-ratio $q=m_1/m_2$, and the length in orbits of the simulation.}
\label{table:etalist4}
\end{table}

%% file: pndescription.tex
Post-Newtonian (PN) theory is a perturbative approach to describing the
motion of a compact object binary, during the slow-motion and weak-field 
regime, i.e. the inspiral phase. The conserved energy of a binary in orbit,
$E$, has been calculated to 3PN order in literature~\citep{Jaranowski:1997ky,
Jaranowski:1999ye,Jaranowski:1999qd,Damour:2001bu,Blanchet:2003gy,
Damour:2000ni,Blanchet:2002mb}.
Using the adiabatic approximation, we treat the course of inspiral as a series
of radially shrinking circular orbits. This is valid during the inspiral when
the angular velocity of the binary evolves more slowly than the orbital 
time-scale. The radial separation shrinks as the binary loses energy to 
gravitational radiation that propagates outwards from the system. 
The energy flux from a binary $F$ is known in PN theory to 3.5PN 
order~\cite{FluxandE3-5PN,Blanchet:2004ek,Blanchet:2005tk,Blanchet:2004bb}.
Combining the energy balance equation, $\D E/\D t = -F$, with Kepler's law 
gives a description of the radial and orbital phase evolution of the binary. 
We start the waveform where the GW frequency enters the sensitive frequency 
band of advanced LIGO, i.e. at $15$Hz. 
Depending on the way the expressions for orbital energy and flux
are combined to obtain the coordinate evolution for the binary,
we get different Taylor\{T1,T2,T3,T4\} time-domain approximants. Using the 
stationary phase approximation~\cite{MatthewsWalker}, frequency-domain 
equivalents of these approximants, i.e. TaylorF$n$, can be constructed. 
Past GW searches have extensively used the TaylorF2 approximant, as it has a
closed form and mitigates the computational cost of generating and numerically
fourier-transforming time-domain template~\cite{Colaboration:2011nz,Abadie:2010yb,
Abbott:2009qj,Abbott:2009tt,Messaritaki:2005wv}.
We refer the reader to Ref.~\cite{PNtheoryLivingReviewBlanchet,JolienGWPhysAst}
for an overview.
From the coordinate evolution, we obtain the emitted gravitational waveform;
approximating it by the quadrupolar multipole $h_{2,\pm 2}$ which is the 
dominant mode of the waveform.

%% file: hybridizationprocedures.tex
The hybridization procedure used for this investigation is described in Sec.~3.3 of Ref.~\cite{MacDonald:2011ne}: The PN waveform, $h_\text{PN}(t)$, is time and phase shifted to match the NR waveform, $h_\text{NR}(t)$, and they are smoothly joined together in a GW frequency interval centered at $\omega_m$ with width $\delta\omega$: 
\begin{equation}\label{eq:omega_match}
\omega_m-\frac{\delta\omega}{2} \le \omega \le \omega_m+\frac{\delta\omega}{2}.
\end{equation}
This translates into a matching interval $t_{\rm min}<t<t_{\rm max}$ because the GW frequency continuously increases during the inspiral of the binary. As argued in Ref.~\cite{MacDonald:2011ne}, we
choose $\delta\omega = 0.1\omega_m$ because it offers a good compromise of suppressing residual oscillations in the matching time, while still allowing $h_\text{PN}(t)$ to be matched as closely as possible to the beginning of $h_\text{NR}(t)$.

The PN waveform depends on a (formal) coalescence time, $t_c$, and phase, $\Phi_c$. These two parameters are determined by minimizing the GW phase difference in the matching interval $[t_\text{min}, t_\text{max}]$ as follows:
\begin{equation}\label{eq:tcphic_bymin}
t_c', \Phi_c' = \mathrm{ \mathop{arg min}_{t_c, \Phi_c}}\int^{t_{\rm max}}_{t_{\rm min}} \big(
  \phi_{\rm PN}(t;t_c,\Phi_c) - \phi_{\rm NR} (t) \big)^2 \rm{d}t,
\end{equation}
where $t_c'$ and $\Phi_c'$ are the time and phase parameters for the best matching between $h_\text{PN}(t)$ and $h_\text{NR}(t)$, and $\phi(t)$ is the phase of the (2,2) mode of the gravitational
radiation. Since we consider only the (2,2) mode, this procedure is identical to time and phase shifting the PN waveform until it has best agreement with NR as measured by the integral in Eq.~(\ref{eq:tcphic_bymin}). The hybrid waveform is then constructed in the form
\begin{equation}
h_\text{H}(t) \equiv \mathcal{F}(t) h_\text{PN}(t;t'_c,\Phi'_c) + \big[1- \mathcal{F}(t)\big]  h_\text{NR} (t), 
\end{equation}
where $\mathcal{F}(t)$ is a blending function defined as
\begin{eqnarray}
\mathcal{F}(t) \equiv 
\left\{
\begin{array}{ll}
  1, &  t < t_{\rm min} \\ 
 \cos^2\frac{\pi(t - t_{\rm min})}{2(t_{\rm max} - t_{\rm min})},\quad\quad &  t_{\rm min}
  \leq t < t_{\rm max} \\ 
  0. & t\geq t_{\rm max}  .
\end{array}
\right.\label{eq:BlendingFunction}
\end{eqnarray}
In this work, we construct all hybrids using the same procedure, Eqs.~(\ref{eq:omega_match})--(\ref{eq:BlendingFunction}), and vary only the PN approximant and the matching frequency $\omega_m$.

%% file: eobnrv2description.tex
Full numerical simulations are available for a limited number of binary 
mass combinations. We use a recently proposed EOB 
model~\cite{BuonannoEOBv2Main}, which we refer to as EOBNRv2, as a substitute
to model the signal from binaries with arbitrary component masses
in this paper. This model was calibrated to most of the numerical simulations
that we use to construct templates banks, which span the range of masses 
we consider here well. So we expect this approximation to hold. We describe the
model briefly here.

The EOB formalism maps the dynamics of a two-body system onto an effective-mass
moving in a deformed Schwarzschild-like background~\citep{EOBOriginalBuonannoDamour}.
The formalism has evolved to use Pad\'{e}-resummations of perturbative 
expansions calculated from PN theory, and allows for the introduction of higher
(unknown) order PN terms that are subsequently calibrated against NR 
simulations of BBHs 
\cite{EOBdevel01,EOBdevel02,EOBNRdevel03,DamourFluxhlm01,EOBNRdevel01}. The EOB 
model proposed recently in Ref.~\citep{BuonannoEOBv2Main} has been calibrated 
to SpEC NR waveforms for binaries of mass-ratios $q=\{1,2,3,4,6\}$, where 
$q\,\equiv \, m_1/m_2$, and is the one that we use in this paper (we will refer
to this model as EOBNRv2).

The dynamics of the binary enters in the metric coefficient of the deformed
Schwarzschild-like background, the EOB Hamiltonian \cite{EOBOriginalBuonannoDamour}, 
and the radiation-reaction force. 
These
are known to 3PN order \cite{EOBOriginalBuonannoDamour,PadeAD} from PN theory.
The 4PN \& 5PN terms were introduced in the potential $A(r)$, which was 
Pad\'{e} resummed and calibrated to NR simulations
\citep{EOBNRdevel01,EOBNRdevel02,EOBNRdevel03,EOBNRdevel04,BuonannoEOBv2Main}.
We use the resummed potential calibrated in Ref.~\cite{BuonannoEOBv2Main} 
(see Eq.~(5-9)). The geodesic dynamics of the reduced mass 
$\mu\,=\,m_1 m_2 / M$ in the deformed background 
is described by
the Hamiltonian $H^{\eff}$ given by Eq.~(3) in \cite{BuonannoEOBv2Main}.
The Hamiltonian describing the conservative dynamics of the binary
(labeled the \textit{real} Hamiltonian $H^{\real}$) is related to 
$\hat{H}^{\eff}$ as in Eq.~(4) of \cite{BuonannoEOBv2Main}.
The inspiral-merger dynamics can be obtained by numerically solving the 
Hamiltonian equations of motion for $H^{\real}$, see e.g. Eq.(10)
of~\cite{BuonannoEOBv2Main}. 

The angular momentum carried away from the binary
by the outwards propagating GWs results in a radiation-reaction force
that causes the orbits to shrink.
This is due to the flux of energy from the binary, which
is obtained by summing over the contribution from each term in the multipolar
decomposition of the inspiral-merger EOB waveform.
Complete resummed expressions for these multipoles~\cite{DamourFluxhlm01} can be 
read off from Eq.(13)-(20) of Ref.~\cite{BuonannoEOBv2Main}. In this paper, 
as for PN waveforms, we model the inspiral-merger part 
by summing over the dominant $h_{2,\pm 2}$ multipoles.

The EOB merger-ringdown part is modeled as a sum of $N$ quasi-normal-modes 
(QNMs),
where $N=8$ for EOBNRv2~\citep{EOBNRdevel01,EOBNRdevel02,EOBNRdevel04,BHRDQNMs}.
The ringdown frequencies depend on the mass and spin of the BH that is formed 
from the coalescence of the binary. The inspiral-merger and ringdown parts are
attached by matching them at the time at which the amplitude of the 
inspiral-merger waveform peaks.
~\citep{EOBNRdevel01,BuonannoEOBv2Main}. The matching procedure followed
is explained in detail Sec.~II~C of Ref.~\citep{BuonannoEOBv2Main}.
By combining them, we obtain the complete waveform for a BBH system.

%% file: mathdefinitions.tex

To assess the recovery of SNR from template banks with NR waveforms or NR-PN 
hybrids as templates, we use the measures proposed in
Ref.~\cite{FittingFactorApostolatos,Sathyaprakash:1991mt,Balasubramanian:1995bm}. 
The gravitational waveform emitted during and driving a BBH coalescence is
denoted as $h(t)$, or simply $h$. The inner product between two 
waveforms $h_1$ and $h_2$ is
\begin{equation}\label{eq:overlap}
(h_1|h_2) \equiv 
4\int^{f_\mathrm{Ny}}_{f_\mathrm{min}}\dfrac{\tilde{h}_1(f)\tilde{h}_2^*(f)}{S_n(f)}\D f,
\end{equation}
where $S_n(f)$ is the one-sided power spectral density (PSD) of the detector
noise, which is assumed to be stationary and Gaussian with zero mean; 
$f_\mathrm{min}$ is the lower frequency cutoff for filtering; $f_\mathrm{Ny}$
is the Nyqyuist frequency corresponding to the waveform sampling rate; and 
$\tilde{h}(f)$ denotes the Fourier transform of $h(t)$.
In this paper, we take $S_n(|f|)$ to be the \textit{zero-detuning high power} 
noise curve for aLIGO, for both bank placement and overlap
calculations~\cite{aLIGONoiseCurve}; and set the lower frequency cutoff 
$f_\mathrm{min} =15$~Hz. The peak GW frequency for the lowest binary masses
that we consider, i.e. for $m_1+m_2\simeq 12M_\odot$, is $\sim 2.1$~kHz during
the ringdown phase. We sample the waveforms at $8192$~Hz, preserving the 
information content up to the Nyquist frequency $f_\mathrm{Ny}=4096$~Hz.
A waveform, h, is normalized (made to be a unit vector) by 
$\hat{h} = h/\sqrt{h | h}$. In addition to being senstive to their 
intrinsic mass parameters, the inner product of two normalized waveforms is 
sensitive to phase and time shift differences between the two, $\phi_{c}$ and
$t_{c}$.  These two parameters ($\phi_c$ and $t_c$) can be analytically
maximized over to obtain the maximized overlap $\Olap$,
\begin{equation}\label{eq:maxnormolap}
\Olap(h_1,h_2) = 
\underset{\phi_c,t_c}{\mathrm{max}}\,\l(\hat{h}_1|\hat{h}_2(\phi_c,t_c)\r),
\end{equation}
which gives a measure of how ``close'' the two waveforms are in the waveform
manifold, disregarding differences in overall amplitude. The \textit{mismatch}
$\mathcal{M}$ between the two waveforms is then
\begin{equation}\label{eq:mismatch}
\mathcal{M}(h_1,h_2) = 1 - \Olap(h_1,h_2).
\end{equation}

Matched-filtering detection searches employ a discrete bank of modeled
waveforms as filters. The optimal signal-to-noise ratio (SNR) is obtained when
the detector strain $s(t)\equiv h^{\tr}(t) + n(t)$ is filtered with the 
\textit{true} waveform $h^{\tr}$ itself, i.e.
\begin{equation}
 \rho_{\mathrm{opt}} = \underset{\phi_c,t_c}{\mathrm{max}}\,\l(h^{\tr}|\hat{h}^{\tr}(\phi_c,t_c)\r) = \leftn h^{\tr}\rightn,
\end{equation}
where $\leftn h^{\tr}\rightn\equiv\sqrt{\left(h^{\tr},h^{\tr}\right)}$ is the
noise weighted norm of the waveform. With a discrete bank of filter templates, 
the SNR we recover
\begin{equation}\label{eq:realoptimalSNR}
 \rho\simeq \Olap(h^{\tr},h_b)\leftn h^{\tr}\rightn = \Olap(h^{\tr},h_b)\,\rho_{\mathrm{opt}},
\end{equation}
where $h_b$ is the filter template in the $b$ank (subscript $b$) that has the
highest overlap with the signal $h^{\tr}$.
The furthest distance to which GW signals can be detected is proportional to 
the matched-filter SNR that the search algorithm finds the signal with. 
Note that $0\leq\Olap(h^{\tr},h_b)\leq 1$, so the recovered SNR
$\rho\leq \rho_{\mathrm{opt}}$ (c.f. Eq.~(\ref{eq:realoptimalSNR})). 
For a BBH population uniformly distributed in spacial volume, the 
detection rate would decrease as $\Olap(h^{\tr},h_b)^3$. Searches that aim at
restricting the loss in the detection rate strictly below 
$10\%\,(\mathrm{or}\, 15\%)$, would require a bank of template waveforms that
have $\Olap$ above $0.965\,(\mathrm{or}\, 0.947)$ with \textit{any} incoming
signal~\citep{WaveformAccuracy2008,WaveformAccuracy2010}.

Any template bank has two sources for loss in SNR: 
(i) the discreteness of the bank grid in the physical parameter space of the 
BBHs, and, (ii) the disagreement between the actual GW signal $h^{\tr}$ and the 
modeled template waveforms used as filters. We de-coupled these to estimate
the SNR loss. Signal waveforms are denoted as $h^\tr_x$ in what follows, 
where the superscript $\tr$ indicates a $tr$ue signal, and the subscript
$x$ indicates the mass parameters of the corresponding binary. Template
waveforms are denoted as $h^\M_b$, where $\M$ denotes the waveform $m$odel, and
$b$ indicates that it is a member of the discrete \textit{b}ank.
\begin{figure}
 \centering
\includegraphics[width=\columnwidth]{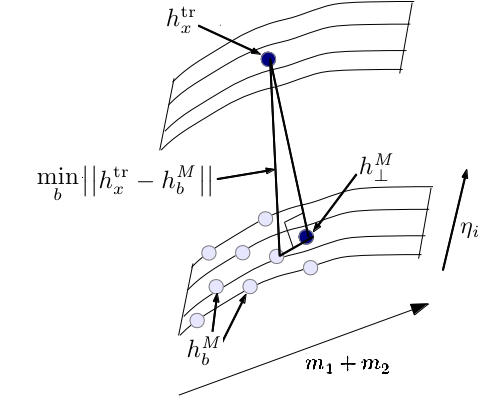}
\caption{We show the \textit{true} (upper) and the \textit{hybrid} (lower) 
waveform manifolds here, with the signal residing in the former, and a discrete
bank of templates placed along lines of constant mass-ratio in the latter. 
Both manifolds are embedded in the same space of all possible waveforms.
The true signal waveform is denoted as $h^{\tr}_x$, while the templates in the
bank are labelled $h^{\M}_b$. The hybrid waveform that matches the signal $H^{\tr}_x$
best is shown as $h^{\M}_\perp$. Also shown is the ``distance'' between
the signal and the hybrid template that has the highest overlap with it.
This figure is qualitatively similar to Fig.~3 of
Ref.~\cite{WaveformAccuracy2008}.}
\label{fig:EFFdiag1}
\end{figure}
Fig.~\ref{fig:EFFdiag1} shows the signal $h^\tr_x$ in its manifold, and the
bank of templates $h^\M_b$ residing in the model waveform manifold, both being
embedded in the same space of all possible waveforms. The 
point $h^\M_\perp$ is the waveform which has the smallest mismatch
in the entire (continuous) model manifold with $h^\tr_x$, i.e.
$h^\M_\perp :\mathcal{M}(h^\tr_x,h^\M_\perp) = \underset{y}{\mn}\,\,\mathcal{M}(h^\tr_x,h^\M_y)$.
The fraction of the optimal SNR recovered at different points $x$ in the
binary mass space can be quantified by measuring the fitting factor $\FF$ of
the bank~\cite{FittingFactorApostolatos},
\begin{equation}\label{eq:ffmismatch}
 \FF(x) = 1 - \underset{b}{\mn}\,\,\mathcal{M}(h^\tr_x, h^\M_b).
\end{equation}
For two waveforms $h_1$ and $h_2$ close to each other in the
waveform manifold: $\leftn h_1\rightn \simeq\leftn h_2\rightn$, and mutually
aligned in phase and time such that the overlap between them is maximized, 
\begin{equation}
  \leftn h_1 - h_2\rightn^2 \simeq 2\l( h_1 |h_1\r)\left(1 - \dfrac{\l( h_1 |h_2\r)}{\sqrt{\l( h_1 |h_1\r)}\sqrt{\l( h_1 |h_1\r)}}\right).
\end{equation}
The mismatch can, hence, be written as 
(c.f. Eq.~(\ref{eq:mismatch}))
\begin{equation}
 \Mis\left(h_1,h_2\right) = \dfrac{1}{2\leftn h_1\rightn^2}\leftn h_1 - 
h_2\rightn^2.
\end{equation}
We note that this equation is an upper bound for Eq.~(25) of
Ref.~\cite{Cannon:2012gq}. 
From this relation, and treating the space embedding the true and model 
waveform manifolds as Euclidean at the scale of template separation, we
can separate out the effects of bank coarseness and template inaccuracies as
\begin{subequations}
\begin{align}
 \FF(x) &= 1 - \underset{b}{\mn}\dfrac{1}{2\leftn h^\tr_x\rightn^2}\leftn h^\tr_x - h^\M_b\rightn^2 ,\\
 &= 1 - \Gamma_\Hyb(x) - \Gamma_\bnk(x)\label{eq:FFGammas};
 \end{align}
\end{subequations}
where 
\begin{equation}
\Gamma_\Hyb(x) \equiv \dfrac{1}{2\leftn h^\tr_x\rightn^2}\leftn h^\tr_x - h^\M_\perp\rightn^2 = \mathcal{M}(h^\tr_x,h^\M_\perp) 
\end{equation}
is the SNR loss from model waveform errors out of the manifold of true signals;
and 
\begin{equation}\label{eq:GammaBank}
\Gamma_\bnk(x) \equiv \underset{b}{\mn}\dfrac{1}{2\leftn h^\tr_x\rightn^2}\leftn h^\M_\perp - h^\M_b\rightn^2 = \underset{b}{\mn}\,\,\mathcal{M}(h^\M_\perp,h^\M_b) 
\end{equation}
is the loss in SNR from the distant spacing of templates in the bank.
The decomposition in Eq.~(\ref{eq:FFGammas}) allows for the measurement of the 
two effects separately. 
NR-PN hybrids have the inspiral portion of the waveform, from PN theory, 
joined to the available late-inspiral and merger portion from NR (as described
in Sec.~\ref{s2:NRpNhybridwaveforms}). Towards the late inspiral, the PN
waveforms accumulate phase errors, contaminating the
hybrids~\cite{MacDonald:2011ne,MacDonald:2012mp}. For each hybrid, we constrain
this effect using mismatches between hybrids constructed from the same NR 
simulation and different PN models, i.e.
\begin{equation}
 \Gamma_\Hyb(x) \leq \mathcal{M}(h^\tr_x,h^\Hyb_x) \lesssim \underset{(i,j)}{\mx}\,\,\mathcal{M}(h^{\M_i}_x,h^{\M_j}_x),
\end{equation}
where  $\M_i = $ TaylorT[1,2,3,4]+NR.
However, this is only possible for a few values of mass-ratio for which NR
simulations are available. We assume $\Gamma_\Hyb$ to be a slowly and smoothly 
varying quantity over the component-mass space at the scale of template grid
separation. At any arbitrary point $x$ in the mass space we approximate 
$\Gamma_\Hyb$ with its value for the ``closest'' template, i.e.
\begin{equation}\label{eq:GammaHybfinal}
 \Gamma_\Hyb(x) \leq \underset{(i,j)}{\mx}\,\,\mathcal{M}(h^{\M_i}_x,h^{\M_j}_x) \simeq \underset{(i,j)}{\mx}\,\,\mathcal{M}(h^{\M_i}_b,h^{\M_j}_b),
\end{equation}
where $h^\M_b$ is the hybrid template in the bank with the highest overlap with 
the signal at $x$. 

The other contribution to SNR loss comes from the discrete placement of 
templates in the mass space. In Fig.~\ref{fig:EFFdiag1}, this is shown in the
manifold of the template model. As NR waveforms (or hybrids) are available
for a few values of mass-ratio, we measure this in the manifold of EOBNRv2
waveforms. The EOBNRv2 model reproduces most of the NR simulations that
were consider here well~\cite{BuonannoEOBv2Main}, allowing for this 
approximation to hold. For the same reason, we expect $h^\EOB_x$ to be close to 
$h^\EOB_\perp$, with an injective mapping between the two. This allows us to 
approximate (c.f. Eq.~(\ref{eq:GammaBank}))
\begin{eqnarray}\label{eq:GammaBankEOB}
\label{eq:Gammabnkfinal}
 \Gamma_\bnk(x) &\simeq & \underset{b}{\mn}\,\,\mathcal{M}(h^\EOB_x,h^\EOB_b).
\end{eqnarray}

In Sec.~\ref{s1:NRonlybank}, we construct template banks that use purely-NR
templates, which have negligible waveform errors. The SNR recovery from such 
banks is characterized with
\begin{equation}\label{eq:NRFFGammas}
 \FF(x) = 1 - \Gamma_\bnk(x),
\end{equation}
where the SNR loss from bank coarseness is obtained using 
Eq.~(\ref{eq:Gammabnkfinal}). In 
Sec.~\ref{s1:NRpNhybridbank},~\ref{s1:futureNRpNhybridbank}, we construct 
template banks aimed at using NR-PN hybrid templates. Their SNR recovery is
characterized using Eq.~(\ref{eq:FFGammas}), where the additional contribution
from the hybrid waveform errors are obtained using Eq.~(\ref{eq:GammaHybfinal}).

%% file: nrbankresults.tex
\begin{figure}
\centering
\includegraphics[width=1.1\columnwidth]{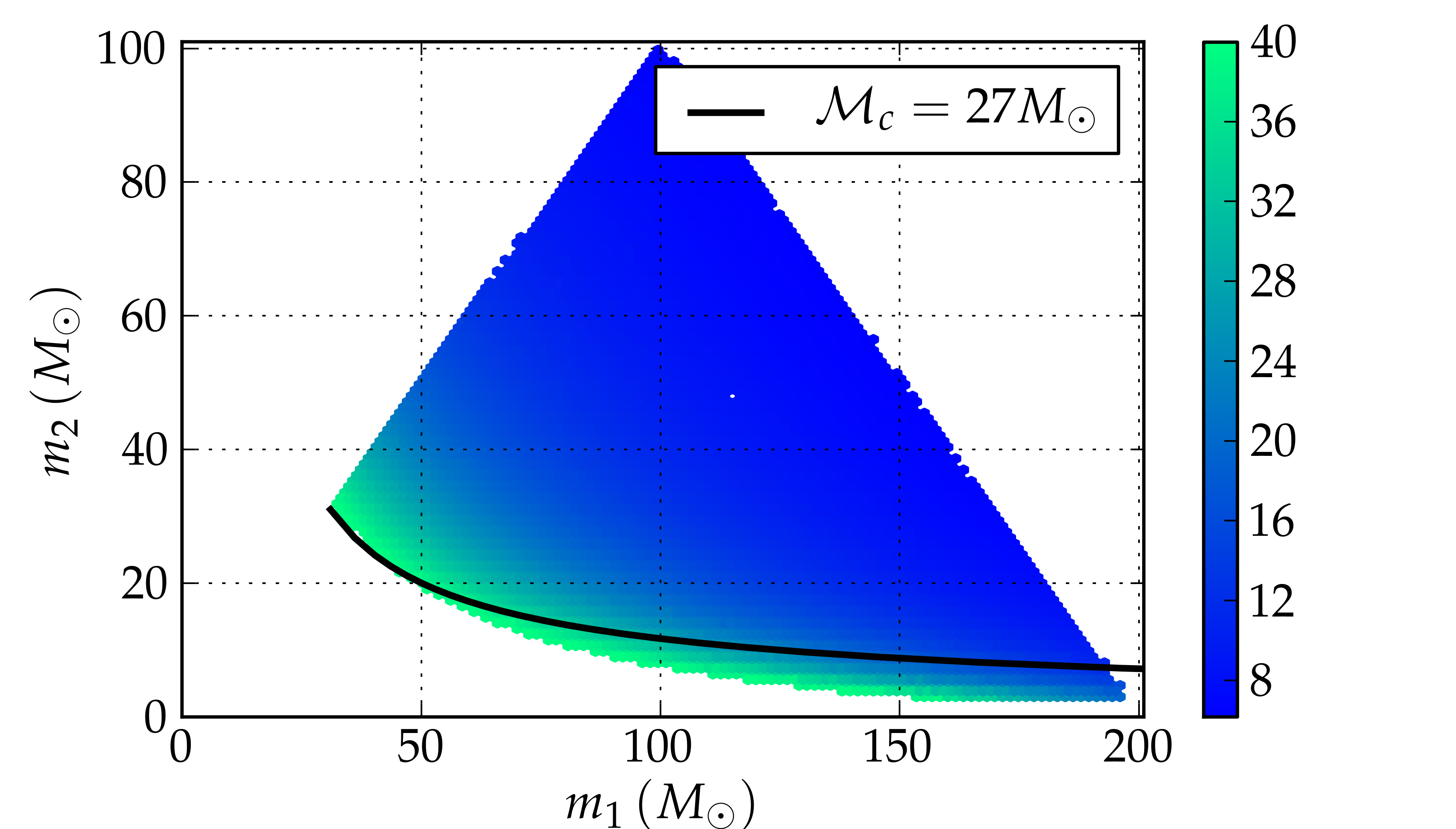}
\caption{The color at each point gives the number 
of waveform cycles $\N_{\cyc}$, for that particular binary, which contain 
$99\%$ of the signal power in the aLIGO sensitivity band. The figure is 
trucated to exclude the region where $\N_{\cyc}>40$. The solid curve shows
the lower bounding edge of the region with $\mathcal{M}_c = 27M_\odot$.}
\label{fig:BBHregion}
\end{figure}

In this section we demonstrate the effectualness of a template bank viable
for using NR waveforms as templates. The gravitational-wave phase of the dominant 
waveform multipole 
extracted from runs at different resolutions was found to converge within 
$\sim 0.3\,\mathrm{rad}$ for $q=3,4,6$, and within $\sim 0.06\,\mathrm{rad}$
for $q=1,2$ at merger (see Fig.~(6) of Ref.~\cite{Buchman:2012dw}, and 
Fig.~(6,7) of Ref.~\cite{BuonannoEOBv2Main} for a compilation). Most of this 
phase disagreement accumulates over a relatively short duration of 
$\sim 50M  - 100M$ before merger, and is significantly lower over the preceding
inspiral and plunge. As the matched-filter SNR accumulates secularly over the
entire waveform, 
these numerical phase errors are negligible in terms of mismatches. We set
$\Gamma_\Hyb = 0$ while computing the fitting factors, so one is left with
considering $\Gamma_\bnk$ to determine the fidelity of the bank (c.f.
Eq.~(\ref{eq:FFGammas})).

With NR simulations as templates, the region that the bank can cover is 
restricted to binaries that have approximately the same number of waveform 
cycles within the sensitive frequency band of the detectors as the simulations
themselves. We take their fiducial length to be $\sim 40$ GW
cycles~\cite{40GWcycles}. For BBHs with 
$3M_{\odot}\leq m_1,m_2\leq 200M_{\odot}$ and $m_1+m_2\leq 200M_{\odot}$ 
we map out the region with $99\%$ of the signal power within $40$ cycles as the
target region of the purely-NR bank. For samples taken over the mass space, we
determine the frequency interval $[f_1,f_2]$ for which
\begin{equation}\label{eq:99percentpower}
 \int_{f_1}^{f_2}\D f \dfrac{|\tilde{h}(f)|^2}{S_n(|f|)} = 
0.99\times\int_{f_\mathrm{min}}^{f_\mathrm{Ny}}\D f \dfrac{|\tilde{h}(f)|^2}{S_n(|f|)}.
\end{equation}
This is done by finding the peak of the integrand in 
Eq.~(\ref{eq:99percentpower}) and integrating symmetrically outwards from 
there, in time, till the interval $[f_1,f_2]$ is found. The number 
of waveform cycles in this interval is
\begin{equation}
 \N_{\cyc} = \dfrac{\Phi( t(f_2) ) - \Phi( t(f_1) )}{2\pi},
\end{equation}
where $\Phi(t)$ is the instantaneous phase of the waveform, 
${h_+(t)\,-\,\ii h_{\times}(t)\,=\,A(t)\,e^{-\ii \Phi(t)}}$, un-wrapped to be a
monotonic function of time. 
We find that for a significant portion of the mass-region, the signal power 
is contained within $40$ waveform cycles. This is shown in 
Fig.~\ref{fig:BBHregion}, where the color at each point gives $\N_{\cyc}$ for
that system, and the region with $\N_{\cyc}> 40$ is excluded. Conservatively, 
this region is bounded by $\mathcal{M}_c = 27M_\odot$, as shown by the solid 
curve in the figure.

We place a bank over this region, using a stochastic method similar 
to Ref.~\citep{Harry:2009ea,Ajith:2012mn,Manca:2009xw}. 
The algorithm begins by taking an empty bank,
corresponding to step $0$. At step $i$, a proposal point $(q,M)$ is picked
by first choosing a value for $q$ from the restricted set
$\mathcal{S}_q=\{1,2,3,4,6,8\}$. The total mass $M$ is subsequently sampled
from the restricted interval corresponding to the pre-drawn $q$. The proposal 
is accepted if the waveform at this point has overlaps $\mathcal{O}< 0.97$ 
with all the templates in the bank from step $i-1$. This gives 
the bank at step $i$. The process is repeated till the fraction of 
proposals being accepted falls below $\sim 10^{-4}$, and the coverage
fraction of the bank is $\gtrsim 99\%$.
To complete the coverage, $100,000$ points are sampled over the region of mass
space depicted in Fig.~\ref{fig:BBHregion}, and $\FF$ of the bank
is computed at each point. With the islands of undercoverage isolated, the points
sampled in these regions are added to the bank, pushing their mass-ratios to 
the two neighboring mass-ratio from $\mathcal{S}_q$. 
This helps accelerate the convergence of the bank, albeit at the cost of 
over-populating it, as the algorithm for computing the $\FF$ for the 
sampled points is parallelizable.
\begin{figure}
\centering
\includegraphics[width=\columnwidth]{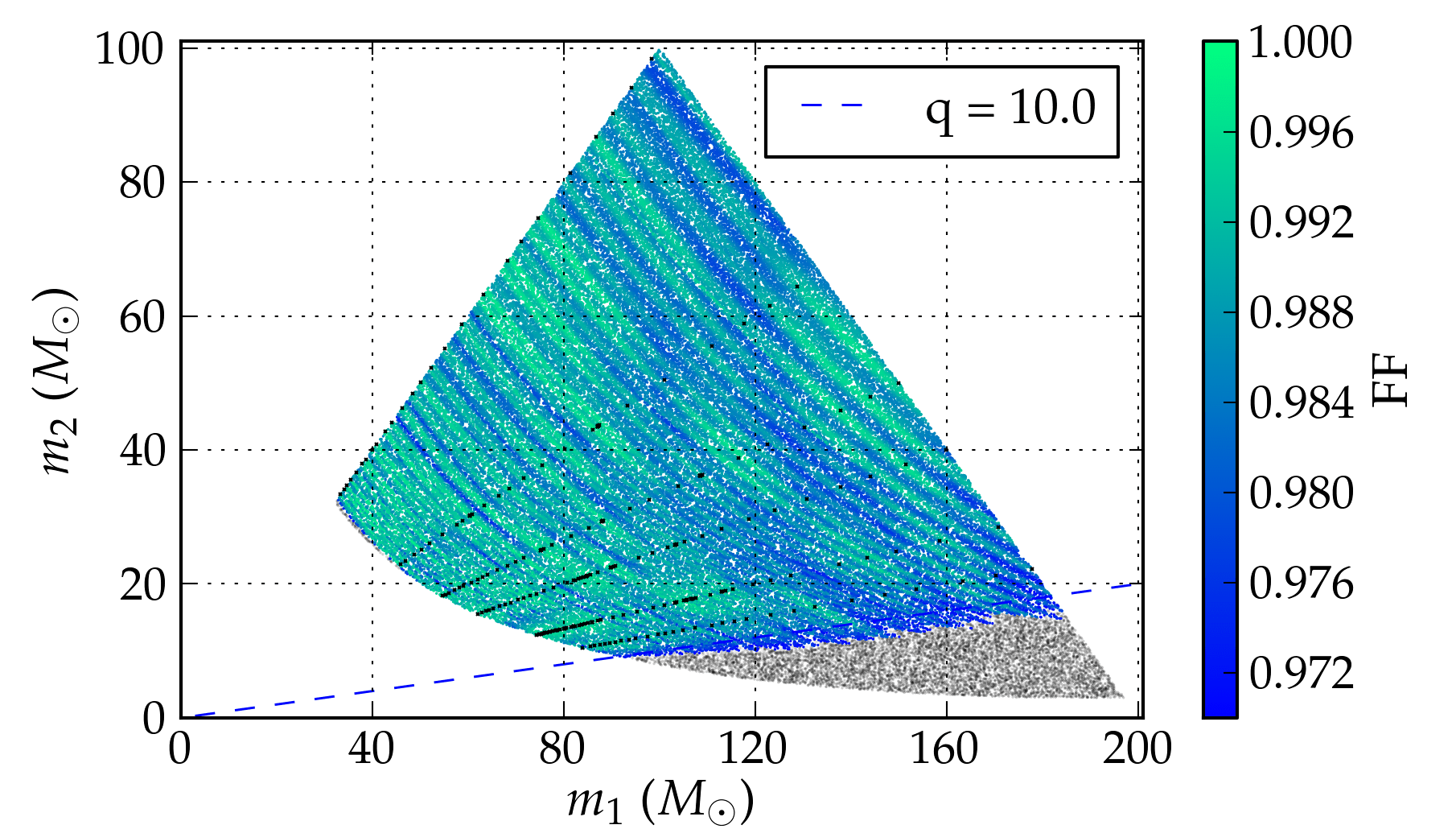}
\caption{The color at each point in the figure gives the
value of $\FF\simeq 1-\Gamma_{\bnk}$ of the bank for that binary, for
the NR bank restricted to $\mathcal{S}_q=\{1,2,3,4,6,8\}$. This is the
same as the fraction of the optimal SNR, for the binary, that the
template bank recovers. The black dots show
the location of the templates in the bank. We note that they all lie
along straight lines of constant $q$ passing through the origin. The region 
shaded light-grey (towards the bottom of the figure) is where the $\FF$ 
drops sharply below $97\%$.}
\label{fig:bank001_01_match}
\end{figure}
\begin{figure}
\centering
\includegraphics[width=\columnwidth]{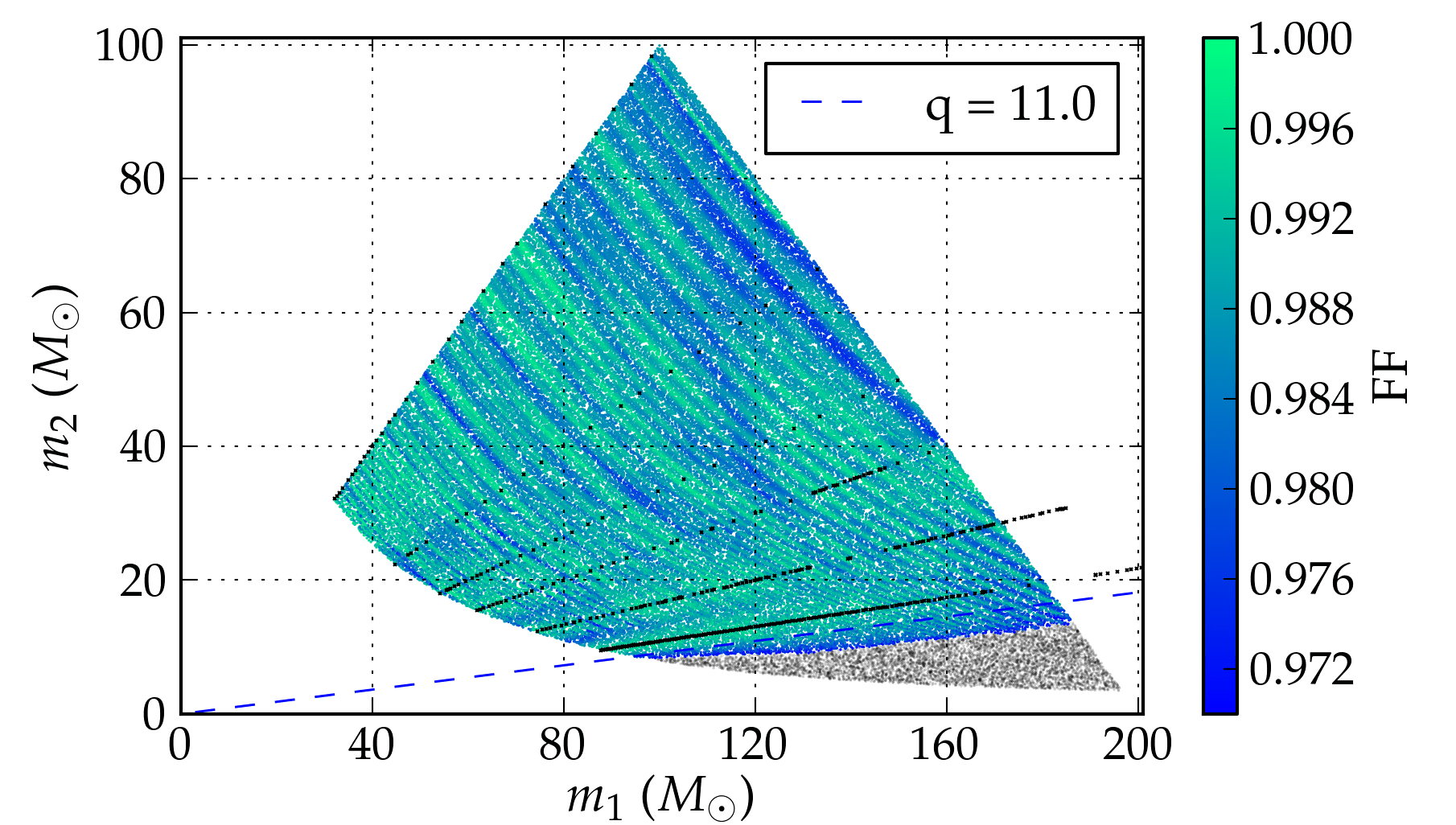}
\caption{This figure is similar to Fig.~\ref{fig:bank001_01_match}.
The color at each point gives the value of 
$\FF\simeq 1-\Gamma_{\bnk}$ of the bank for that binary, for
the NR bank restricted to $\mathcal{S}_q=\{1,2,3,4,6,9.2\}$.
The black dots show the location of the templates in the bank. 
The region shaded light-grey (towards the bottom of the figure) is 
where the $\FF$ drops below $97\%$. We note that with an additional
NR waveform for mass ratio $q=9.2$, the coverage of the bank is 
extended to include binaries with $10\leq q\leq 11$.}
\label{fig:bank006_01_match}
\end{figure}

We asses the effectualness of the bank, as discussed in 
Sec.~\ref{s1:quantifyingerrors}, using Eq.~(\ref{eq:NRFFGammas}).
We draw a population of $100,000$ BBH signals, uniformly from the binary 
mass space, and filter them through the bank. Fig.~\ref{fig:bank001_01_match}
shows the $\FF$, or the fraction of the optimal SNR recovered by the bank. 
The region shown is restricted to binaries with $N_{\cyc}\leq 40$.
The black dots in the figure show the position of templates in the bank. 
The bank recovers $\geq 97\%$ of the optimal SNR over the entire region 
of interest for $q\leq 10$. We 
propose an additional simulation for $q=9.2$, to increase the coverage to
higher mass-ratios. Substituting this for $q=8$ in the set of allowed 
mass-ratios $\mathcal{S}_q$, we place another bank as before, with
$\mathcal{S}_q=\{1,2,3,4,6,9.2\}$. The SNR loss from this bank is shown in
Fig.~\ref{fig:bank006_01_match}. This bank recovers $\geq 97\%$ of the SNR for
systems with $q\leq 11$ and $\N_{\cyc}\leq 40$. The choice of the additional
simulation at $q=9.2$ was made by choosing a value \textit{close-to} the 
highest possible value of $q$ that does not lead to under-coverage in the 
region between $q=6$ and that value. The \textit{exact} highest allowed 
value was not chosen to reduce the sensitivity of the coverage of the bank
to fluctuations in detector sensitivity.

We conclude that with only six NR waveforms for non-spinning BBHs, that are
$\sim 20$ orbits (or $40$ GW cycles) in length, a template bank can be
constructed that is effectual for detecting binaries with chirp mass above
$27M_\odot$ and $1\leq q\leq 10$. With an additional 
simulation for $q=9.2$, this bank can 
be extended to higher mass-ratios, i.e. to $1\leq q\leq 11$.

%% file: hybridbankresults.tex
The template bank contructed in Sec.~\ref{s1:NRonlybank} is effectual for 
GW detection searches focussed at relatively massive binaries with 
$\mathcal{M}_c \gtrsim 27M_\odot$. As the NR waveforms are restricted to a
small number of orbits, it is useful to consider NR-PN hybrids to bring the
lower mass limit down on the template bank. PN waveforms can be generated
for an arbitrarily large number of inspiral orbits, reasonably accurately and
relatively cheaply. Thus, a hybrid waveform comprised of a long PN early-inspiral 
and an NR late-inspiral, merger, and ringdown could also be arbitrarily long. 
There are, however, uncertainties in the PN waveforms, due to
the unknown higher-order terms. During the late-inspiral and merger phase,
these terms become more important and the PN description becomes
less accurate. In addition, when more of the late-inspiral is
in the detector's sensitivity frequency range, hybrid waveform mismatches 
due to the PN errors become increasingly large, and reduce the
recovered SNR. Thus, when
hybridizing PN and NR waveforms, there must be enough NR orbits that the PN
error is sufficiently low for the considered detector noise-curve. In 
this section, we construct an NR + PN hybrid template bank, for currently
available NR waveforms, and determine the lowest value of binary masses to 
which it covers.

\begin{figure}
\includegraphics[width=0.9\columnwidth, trim=20 17 75 75]{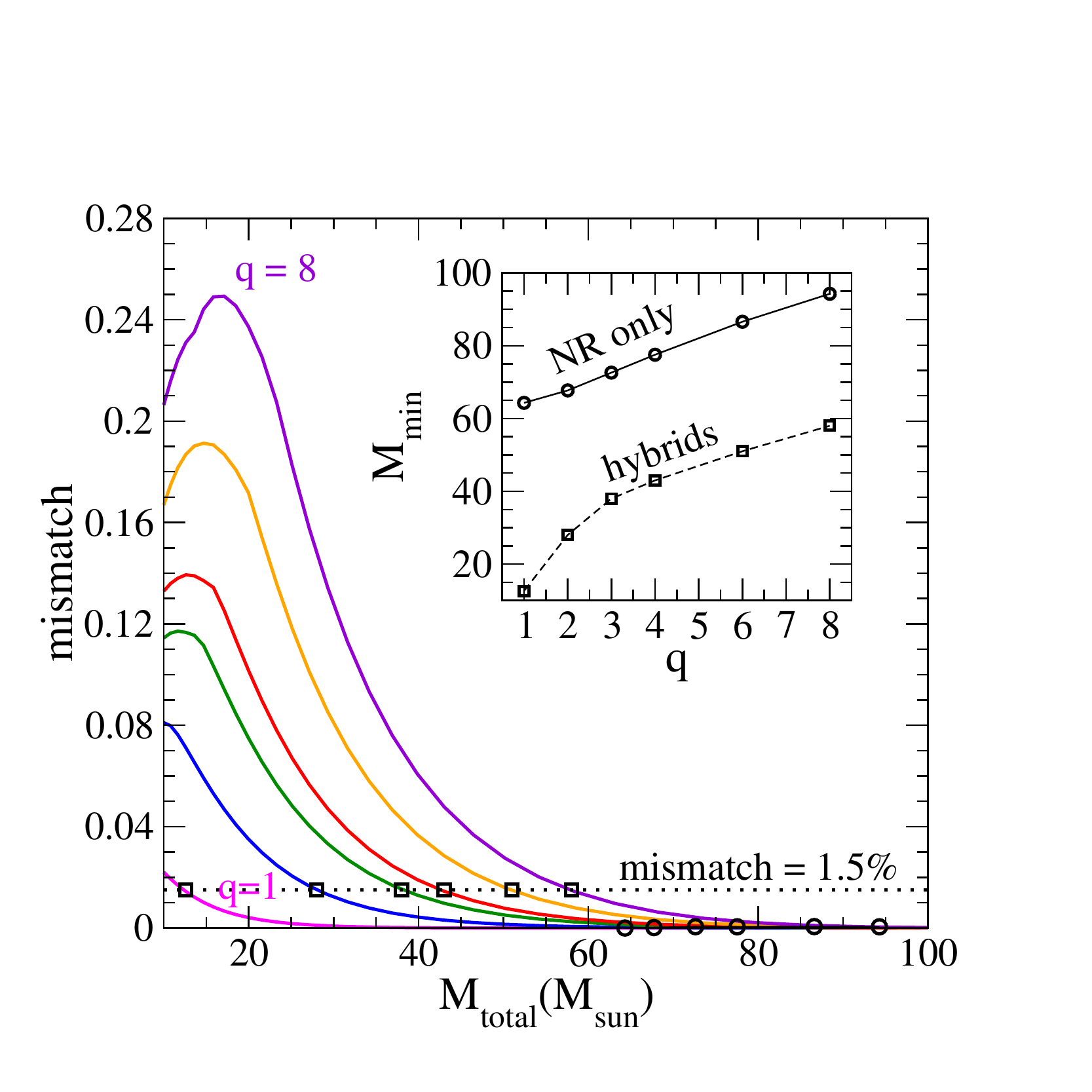}
\caption{\label{fig:Current-NR-PN-Errors}Bounds on mismatches of PN-NR
  hybrid waveforms, for the currently existing NR simulations. The PN
  error is for hybrids matched at $M\omega_m=0.025$ for $q=1$,
  $M\omega_m=0.038$ for $q=2$, and $M\omega_m=0.042$ for
  $q=3,4,6,8$. The black circles indicate the lower bound of the
  template bank in Sec.~\ref{s1:NRonlybank}. The black square show the
  lower bound with a hybrid error of 1.5\%. The inset shows these
  lower bounds as a function of mass ratio.} 
\end{figure}

The hybrids we use are constructed by joining the PN and NR portions, as 
described in Sec.~\ref{s2:NRpNhybridwaveforms}. The number of orbits before 
merger at which they are joined depends on the length of the available NR 
waveforms. We estimate the PN waveform errors using hybridization
mismatches $\Gamma_\Hyb$, as discussed in Sec.~\ref{s1:quantifyingerrors}. 
Fig.~\ref{fig:Current-NR-PN-Errors} shows the same for all the hybrids, as a 
function of total mass. In terms of orbital frequency, these are
matched at $M\omega_m=0.025$ for $q=1$, $M\omega_m=0.038$ for $q=2$,
and $M\omega_m=0.042$ for $q=3,4,6,8$. In terms of number of orbits
before merger, this is 31.9 orbits for $q=1$, 17.8 orbits for $q=2$,
16.9 orbits for $q=3$, 18.4 orbits for $q=4$, 21.6 orbits for $q=6$,
and 25.1 orbits for $q=8$. The dotted line indicates a mismatch of
$1.5\%$, a comparatively tight bound that leaves flexibility to accommodate
errors due to template bank discreteness. The black circles show the hybrid
mismatches at the lower mass bound of the NR-only template bank in
Sec.~\ref{s1:NRonlybank}, which are negligible. The inset shows this minimum 
mass as a function of mass ratio, as well as the minimum attainable mass if we
accept a hybrid error of $1.5\%$. At lower masses, the mismatches increase
sharply with more of the PN part moving into the Advanced LIGO sensitivity band.
This is due to the nature of the frequency dependence of the detector 
sensitivity. The detectors will be relatively very sensitive to a relatively 
short frequency band. As the hybridization frequency sweeps through that band, 
the hybrid errors rise sharply. They fall again at the lowest masses, for which
mostly the PN portion stays within the sensitive band.

\begin{figure*}
\begin{center}
\includegraphics[width=\columnwidth]{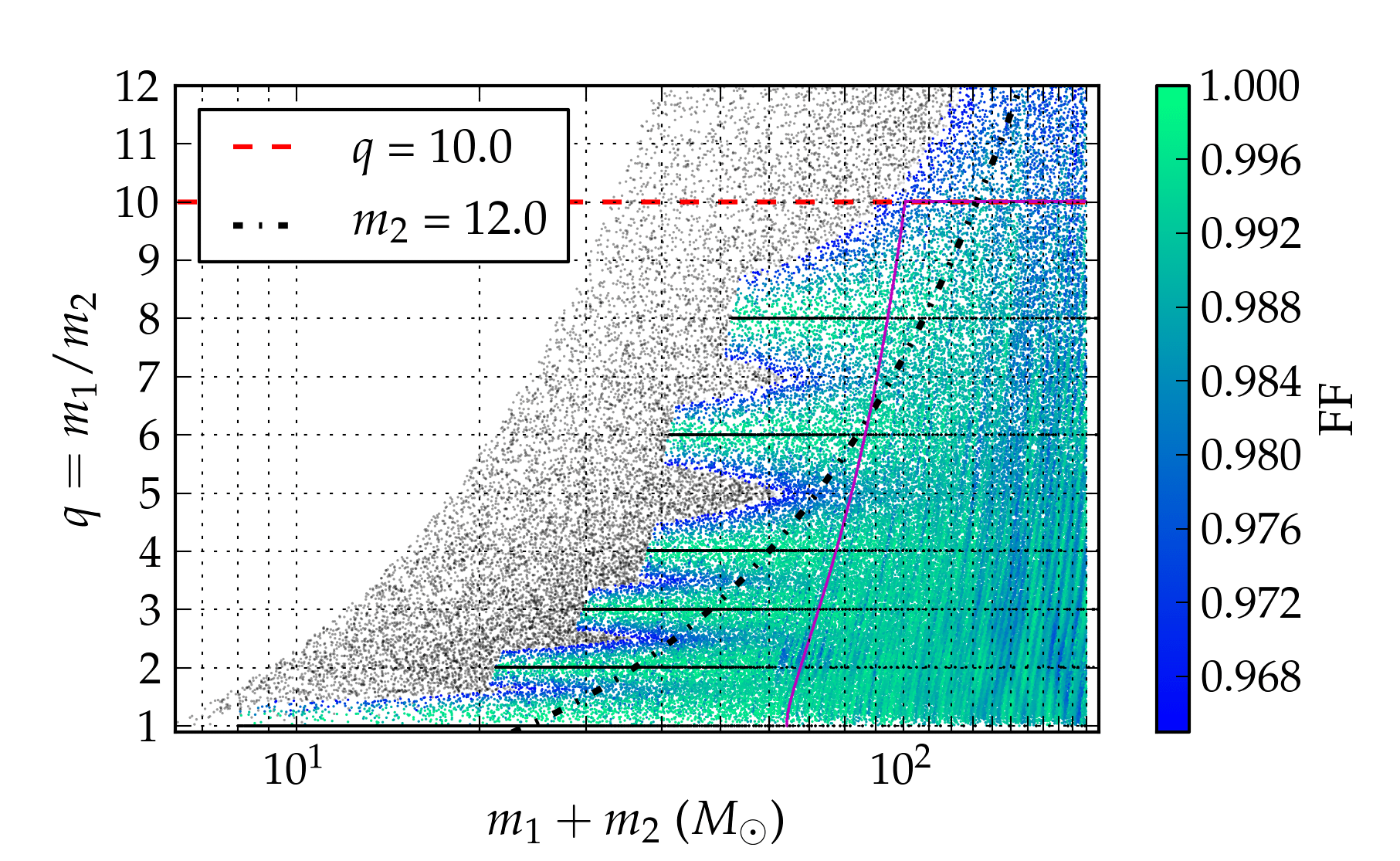}
\includegraphics[width=\columnwidth]{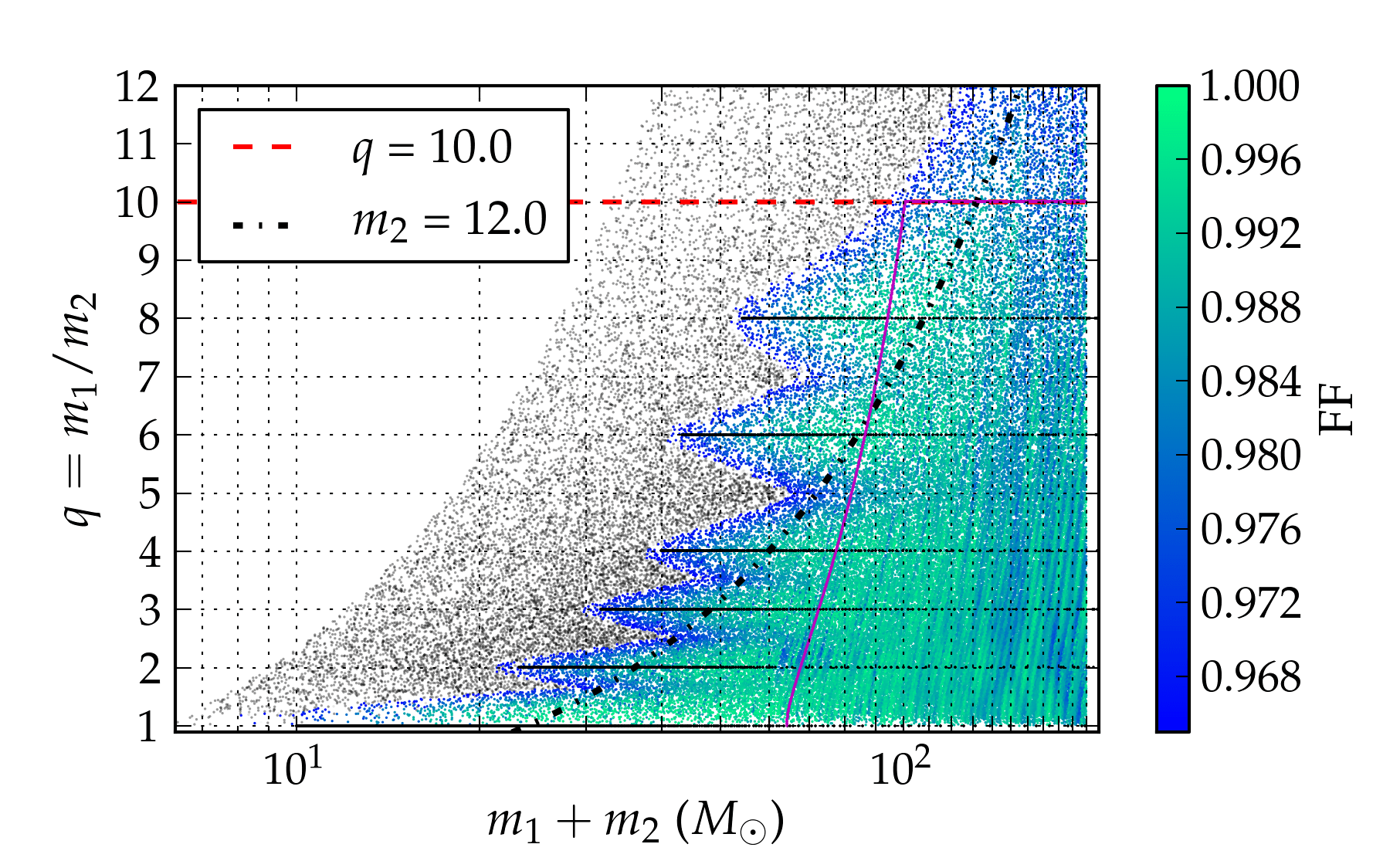}
\caption{\label{fig:Current-hybrids-stochastic-FF}These figures show fitting
  factors $\FF$ obtained when using a discrete mass-ratio template bank for
  $q=1,2,3,4,6,8$. For each mass-ratio, the templates are extended down 
  to a total mass where the NR-PN hybridization mismatch becomes
  $3\%$. The bank is placed using the stochastic algorithm, similar to 
  Ref.~\cite{Harry:2009ea,Ajith:2012mn,Manca:2009xw}. 
  The black dots show the location
  of the templates. The fitting factor on the left plot does 
  {\em not} take into account the hybridization error, and therefore shows the
  effect of the sparse placement of the templates alone. 
  The right plot accounts for the hybridization error
  and gives the actual fraction of the optimal SNR that would be recovered
  with this bank of NR-PN hybrid templates. The region bounded by the magenta 
  (solid) line in both plots indicates the lower end of the coverage of the 
  bank of un-hybridized NR waveforms. Lastly, the shaded grey dots show the 
  points where the fitting factor was below $96.5\%$.}
\end{center}
\end{figure*}
\begin{figure*}
\begin{center}
\includegraphics[width=\columnwidth]{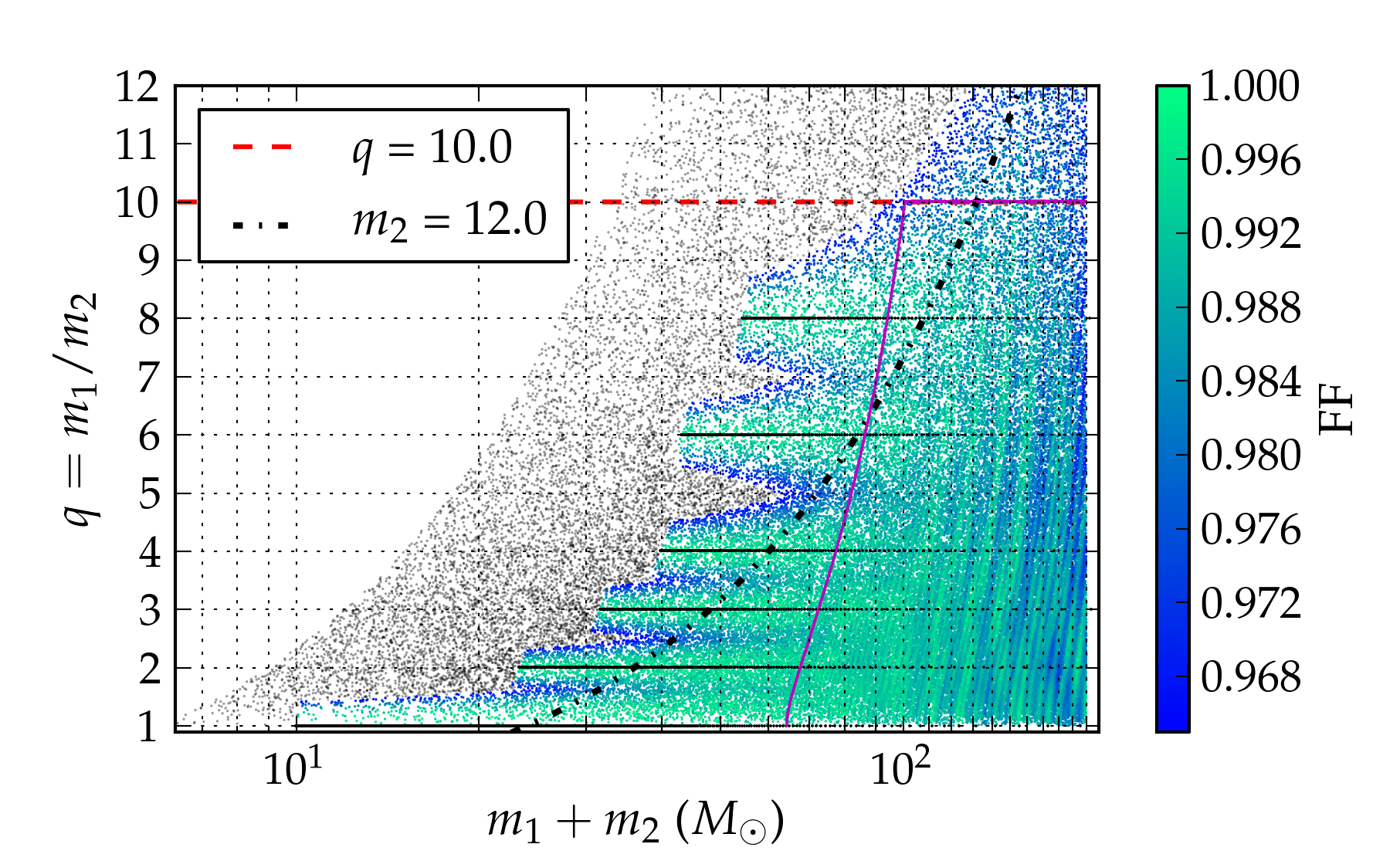}
\includegraphics[width=\columnwidth]{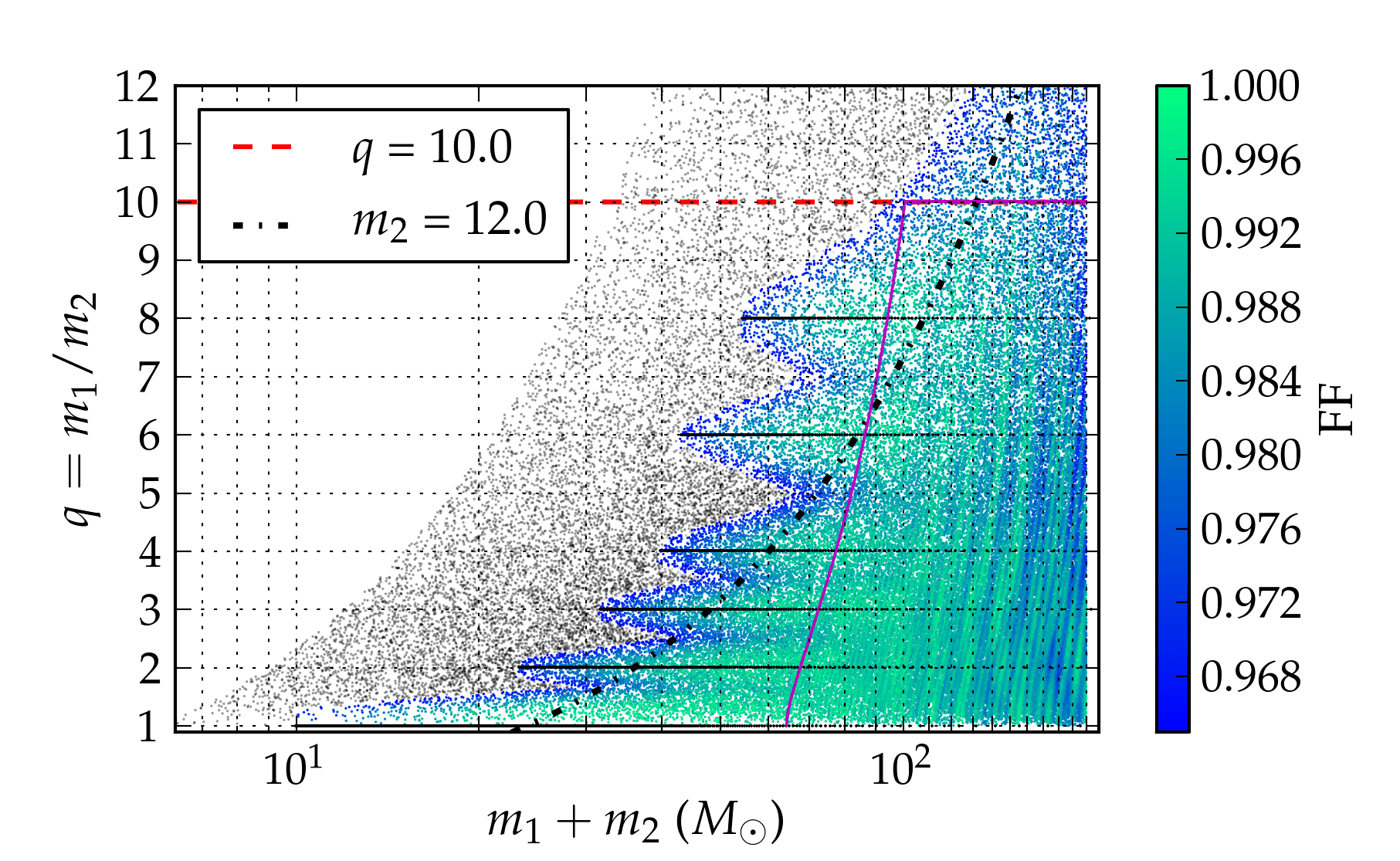}
\caption{\label{fig:Current-hybrids-FF}These figures are similar to 
  Fig.~\ref{fig:Current-hybrids-stochastic-FF}. The figures show fitting
  factors $\FF$ obtained when using a discrete mass-ratio template bank for
  $q=1,2,3,4,6,8$. For each mass-ratio, the templates are extended down 
  to a total mass where the NR-PN hybridization mismatch becomes
  $3\%$. Templates are placed independently for each mass-ratio, and span the 
  full range of total masses. For each mass-ratio, neighboring templates are 
  required to have an overlap of $97\%$. The union of the six single-$q$ 
  one-dimensional banks is taken as the final bank. The black dots show the 
  location of the templates. The fitting factor on the left plot does 
  {\em not} take into account the hybridization error, and therefore shows the
  effect of the sparse placement of the templates alone. The right plot accounts
  for the hybridization error
  and gives the actual fraction of the optimal SNR that would be recovered
  with this bank of NR-PN hybrid templates. The region bounded by the magenta 
  (solid) line in both plots indicates the lower end of the coverage of the 
  bank of un-hybridized NR waveforms. Lastly, the shaded grey dots show the 
  points where the fitting factor was below $96.5\%$.}
\end{center}
\end{figure*}

\begin{figure*}
\begin{center}
\includegraphics[width=\columnwidth]{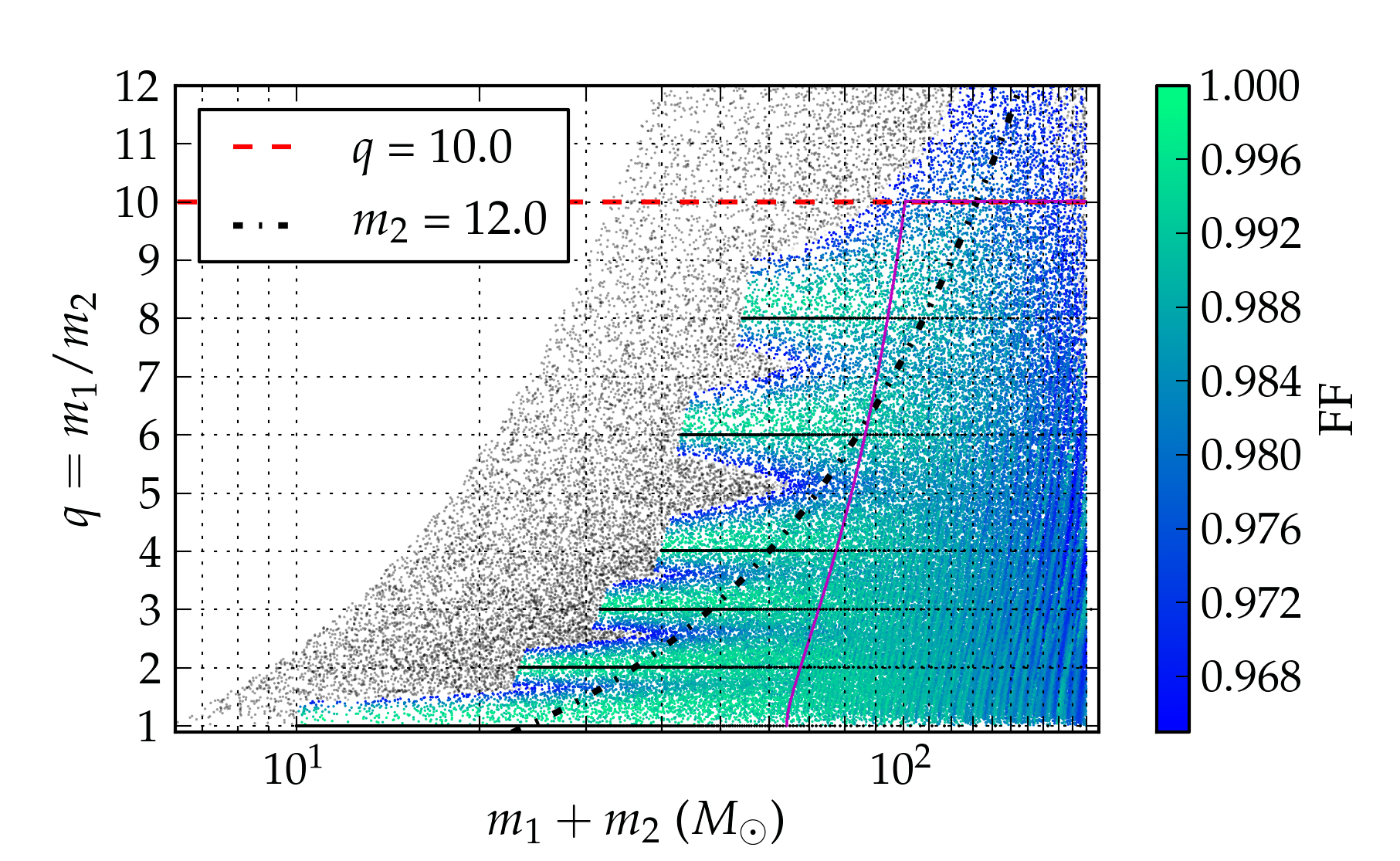}
\includegraphics[width=\columnwidth]{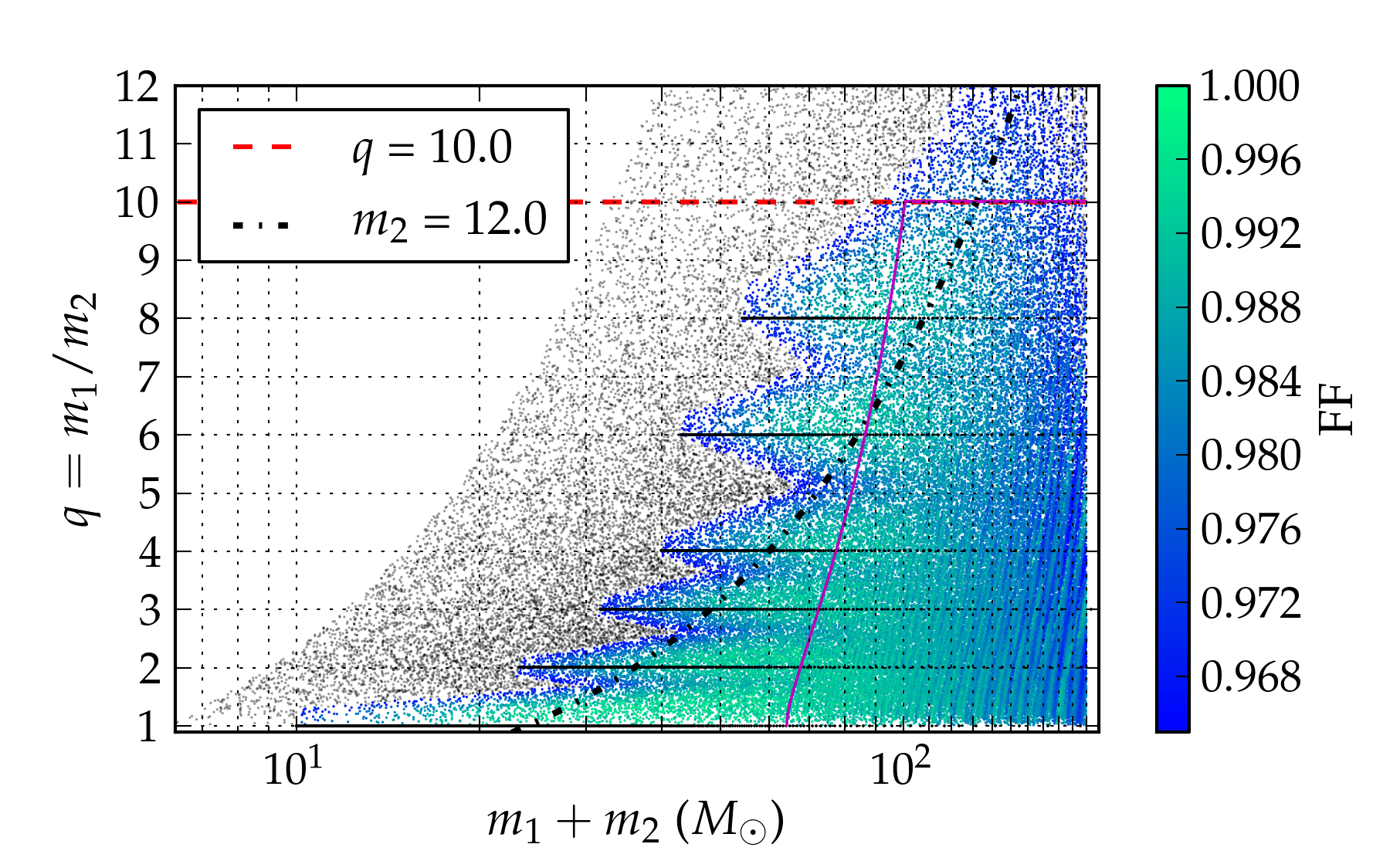}
\caption{\label{fig:Current-real-hybrids-FF}This figure is similar to 
  Fig.~\ref{fig:Current-hybrids-FF}. The figures show fitting
  factors $\FF$ obtained when using a discrete mass-ratio template bank for
  $q=1,2,3,4,6,8$. Templates are placed independently for each mass-ratio, and 
  span the range of total masses, down to the region where the hybrid errors
  become $3\%$. For each mass-ratio, neighboring templates are 
  required to have an overlap of $97\%$. The union of the six single-$q$ 
  one-dimensional banks is taken as the final bank. The black dots show the 
  location of the templates. The GW signals are modeled using the EOBNRv2
  approximant~\cite{BuonannoEOBv2Main}, while TaylorT4+NR hybrids are used as
  templates. The fitting factor on the left plot shows the combined effect of 
  the sparse placement of the templates, and the (relatively small) 
  disagreement between the hybrid and EOBNRv2 waveforms. The right plot
  explicitly accounts for the hybridization error and gives the (conservative)
  actual fraction of the optimal SNR that would be recovered
  with this bank of NR-PN hybrid templates. The region bounded by the magenta 
  (solid) line in both plots indicates the lower end of the coverage of the 
  bank of un-hybridized NR waveforms. Lastly, the shaded grey dots show the 
  points where the fitting factor was below $96.5\%$.  }
\end{center}
\end{figure*}

We now consider template banks viable for hybrids constructed from currently 
available NR waveforms at mass ratios $q=1,2,3,4,6,8$. The lower mass limit,
in this case, is extended down to masses where the hybridization error 
exceeds $3\%$. We demonstrate two independent methods of laying
down the bank grid. First, we use the stochastic placement method that proceeds
as described in Sec.~\ref{s1:NRonlybank}. The templates are sampled over the
total mass - mass-rato $(M,q)$ coordinates, sampling $q$ from the restricted
set. The total mass $M$ is sampled from the continuous interval between the 
lower mass limit, which is different for each $q$, and the upper limit of
$200M_\odot$. To assess the SNR loss from the sparse placement
of the templates, we simulate a population of $100,000$ BBH signal waveforms,
with masses sampled with $3M_\odot\leq m_{1,2}\leq 200M_\odot$ and 
$M\leq 200M_\odot$, and filter them through the bank. This portion of the SNR
loss needs to be measured with both signals and templates in the same waveform
manifold. We use the EOBNRv2 approximant~\cite{BuonannoEOBv2Main} to model both, 
as it has been calibrated to most of the NR waveforms we consider here, and it
allows us to model waveforms for arbitrary systems. The left panel of
Fig.~\ref{fig:Current-hybrids-stochastic-FF} shows the
fraction of the optimal SNR that the bank recovers, accounting for its 
discreteness alone. We observe that, with just six mass-ratios, the bank 
can be extended to much lower masses before it is limited by the restricted
sampling of mass-ratios for the templates. For binaries with both black-holes 
more massive than $\sim 12M_\odot$, the spacing between mass-ratios was found 
to be sufficiently dense. The total SNR loss, after subtracting out 
the hybrid mismatches from Fig.~\ref{fig:Current-NR-PN-Errors}, are shown in 
the right panel of Fig.~\ref{fig:Current-hybrids-stochastic-FF}. 
At the lowest masses, the coverage shrinks between the lines of constant $q$ 
over which the templates are placed, due to the hybrid errors increasing 
sharply. We conclude that this bank is viable for hybrid templates for GW 
searches for BBHs with $m_{1,2}\geq 12M_\odot$, $1\leq q\leq 10$, and 
$M\leq 200M_\odot$. Over this region the bank will recover more than $96.5\%$
of the optimal SNR. This is a significant increase over the coverage allowed 
for with the purely-NR bank, the region of coverage of which is shown in the 
right panel of Fig.~\ref{fig:Current-hybrids-FF}, bounded at lowest masses by 
the magenta (solid) curve.

Second, we demonstrate a non-stochastic algorithm of bank placement, with 
comparable results. We first construct six independent bank grids, each
restricted to one of the mass-ratios $q=1,2,3,4,6,8$, and spanning the full 
range of total masses. The spacing between neighboring templates is
given by requiring that the overlap between them be $97\%$. 
We take the union of these banks as the final 
two-dimensional bank. As before, we measure the SNR loss due to discreteness of
the bank and the waveform errors in the templates separately. To estimate the
former, we simulate a population of $100,000$ BBH systems, and filter
them through the bank. The signals and the templates are both modeled
with the EOBNRv2 model. The left panel of Fig.~\ref{fig:Current-hybrids-FF}
reveals the fraction of SNR recovered over the mass space, accounting for the 
sparsity of the bank alone, i.e. $1-\Gamma_\mathrm{bank}$. At lower masses, 
we again start to see gaps between the lines of constant mass ratio which become
significant at $m_{1,2} \leq 12M_\odot$. 
The right panel of Fig.~\ref{fig:Current-hybrids-FF} shows the final fraction 
of the optimal SNR recovered, i.e. the $\FF$ as defined in Eq.~(\ref{eq:FFGammas}). 
As before, these are computed by subtracting out the hybrid mismatches 
$\Gamma_\Hyb$ in addition to the discrete mismatches, as described in
Sec.~\ref{s1:quantifyingerrors}. 

The efficacy of both methods of template bank construction
can be compared from Fig.~\ref{fig:Current-hybrids-stochastic-FF} and 
Fig.~\ref{fig:Current-hybrids-FF}. We observe that the final banks from either
of the algorithms have very similar SNR recovery, and are both effectual over
the range of masses we consider here. Both were also found to give a very 
similar number of templates. The uniform-in-overlap method yields a grid 
with $2,325$ templates. The stochastic bank, on the other hand, was placed with a
requirement of $98\%$ minimal mismatch, and had $2,457$ templates. 
This however includes templates with $m_{1,2} < 12M_\odot$. Restricted
to provide coverage over the region with $m_{1,2}\geq 12M_\odot$, $1\leq q\leq 10$,
and $M\leq 200M_\odot$, the two methods yield banks with $627$ and $667$
templates respectively. The size of these banks is comparable to one 
constructed using the second-order post-Newtonian TaylorF2 hexagonal 
template placement method~\cite{SathyaBankPlacementTauN,BabaketalBankPlacement,
SathyaMetric2PN,Cokelaer:2007kx}, 
which yields a grid of $522$ and $736$ templates, for a minimal
match of $97\%$ and $98\%$, respectively.

Finally, we test the robustness of these results using TaylorT4+NR hybrids 
as templates. As before, we simulate a population of $100,000$ BBH signal 
waveforms. As we do not have hybrids for arbitary binary
masses, we model the signals as EOBNRv2 waveforms. This population is filtered
against a bank of hybrid templates. The SNR recovered is shown in the left 
panel of Fig.~\ref{fig:Current-real-hybrids-FF}. Comparing with the left panels 
of Fig.~\ref{fig:Current-hybrids-stochastic-FF},~\ref{fig:Current-hybrids-FF}, 
we find that the EOBNRv2 manifold is a reasonable approximation for the hybrid 
manifold; and that, at lower masses, there is a small systematic bias in the 
hybrids towards EOBNRv2 signals with slightly higher mass-ratios.
The right panel of Fig.~\ref{fig:Current-real-hybrids-FF} 
shows the fraction of optimal SNR recovered after subtracting out the hybrid
mismatches from the left panel. The similarity of the $\FF$ distribution between 
the right panels of Fig.~\ref{fig:Current-real-hybrids-FF} and 
Fig.~\ref{fig:Current-hybrids-stochastic-FF},~\ref{fig:Current-hybrids-FF}
is remarkable. This gives us confidence that the EOBNRv2 model is a good 
approximation for testing NR/hybrid template banks, as we do in this paper; 
and that a template bank of NR+PN hybrids is indeed effectual for binaries 
with $m_{1,2}\geq 12M_\odot$, $M\leq 200M_\odot$ and $q\leq 10$.

%% file: futurehybridbankresults.tex
\begin{figure*}
\begin{center}
\includegraphics[width=0.33\textwidth, trim=17 20 75 40]{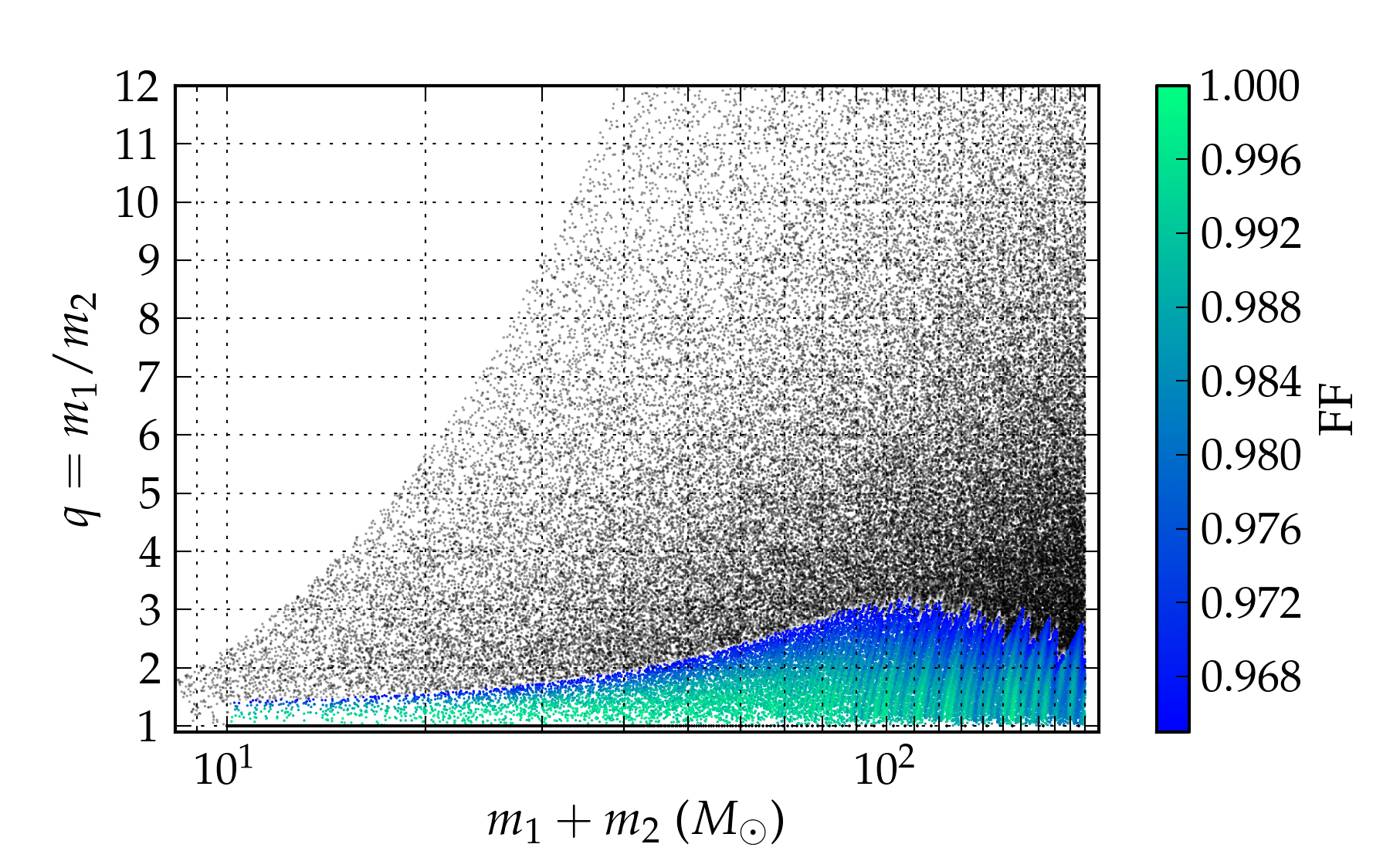}
\includegraphics[width=0.33\textwidth, trim=17 20 75 40]{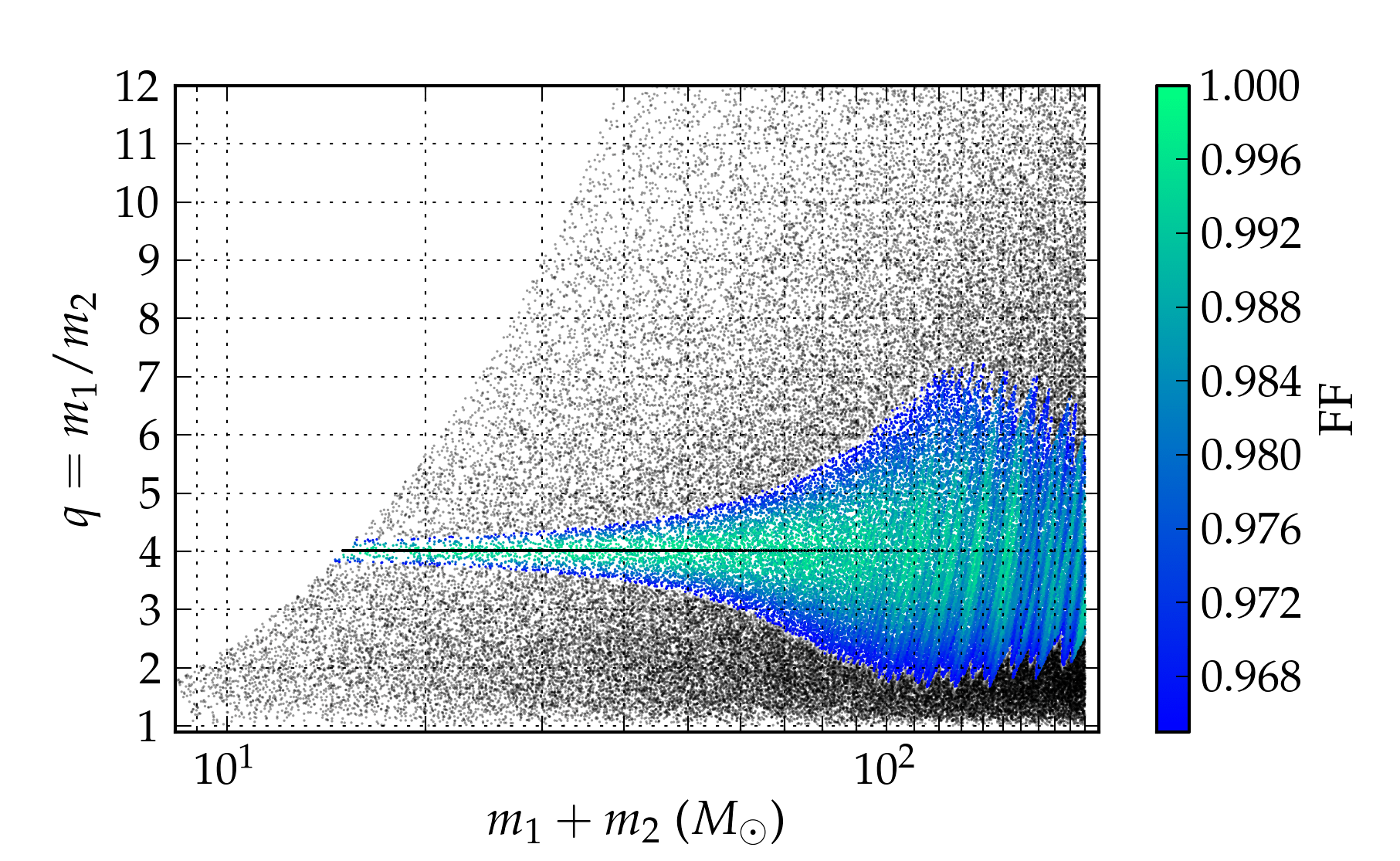}
\includegraphics[width=0.33\textwidth, trim=17 20 75 40]{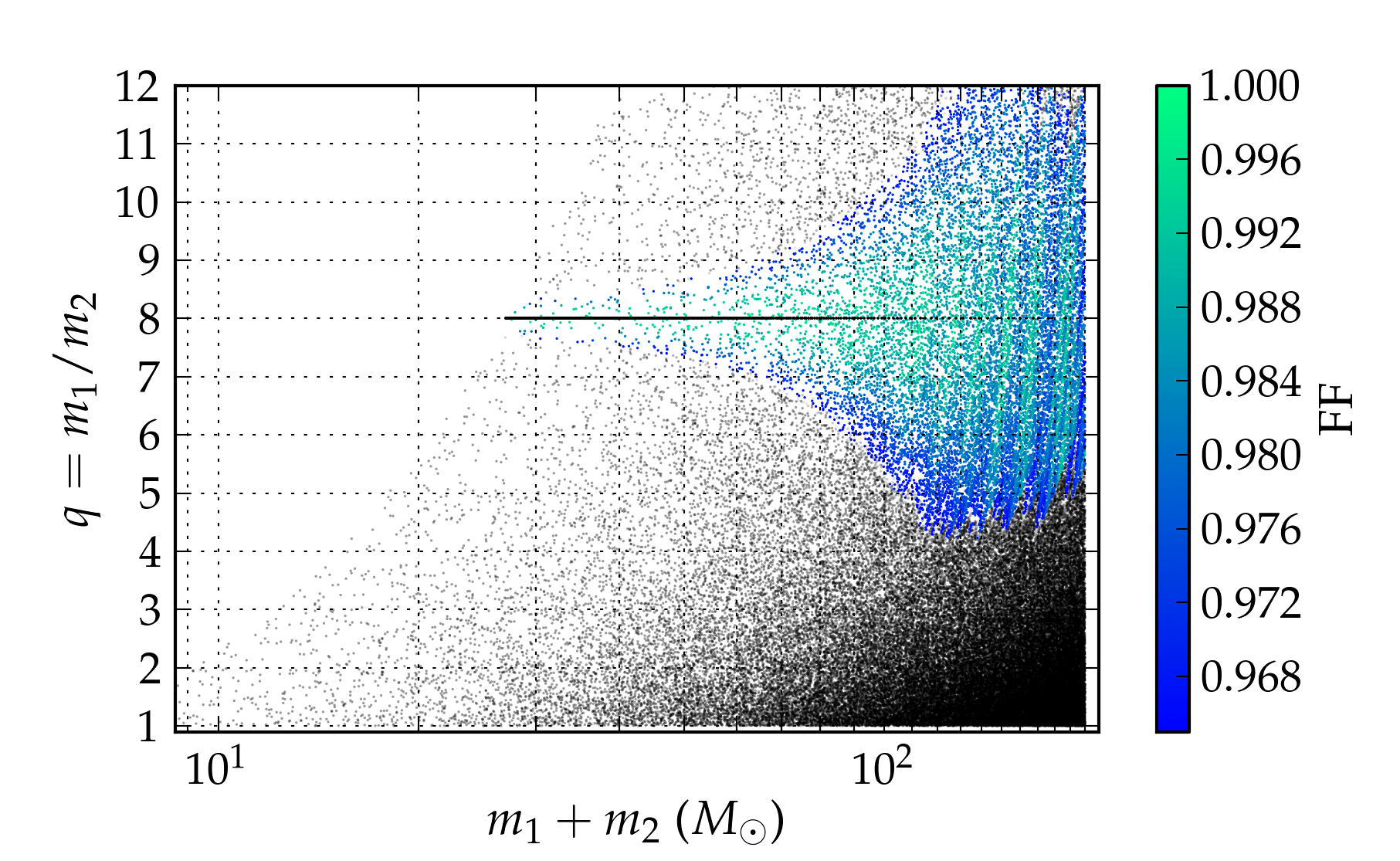}
\caption{\label{fig:separate_q148} These figures show the coverage of template
  banks restricted to single mass-ratios, i.e. (from left to right) 
  $q = 1, 4, 8$. We note that at 
  higher total masses, the templates are correlated to simulated signals for
  considerably different mass-ratios, than at lower total masses. This agrees
  with what we expect as with decreasing total mass, the number of cycles in
  the sensitive frequency band of Advanced LIGO increases.} 
\end{center}
\end{figure*}

The last sections outlined properties of template banks of NR
waveforms (and their hybrids) which are available today. 
We also investigate the parameter and length requirements for future NR 
simulations, that would let us contruct detection template banks all the
way to $M=m_1+m_2=12M_\odot$. This lower limit was chosen following 
Ref.~\cite{Brown:2012nn,CompTemplates2009} which showed that the region with
$M\lesssim 12M_\odot$ can be covered with banks of post-Newtonian inspiral-only
waveforms.

Constructing such a bank is a two-step process.  First, we pick mass-ratios 
that allow construction of such a bank given long enough waveforms for these
mass-ratios. Second, one needs to determine the necessary length of the NR 
portion of the waveforms, such that the PN-hybridization error is sufficiently
low for all masses of interest.

To motivate the first step, Fig.~\ref{fig:separate_q148} shows the coverage of
banks that sample from a {\em single} mass-ratio each (from left to right: $q=1,4,8$). We see that the resolution
in $q$ required at lower values of $M$ increases sharply below 
$M\sim 60M_\odot$. This follows from the increase in the number of waveform
cycles in aLIGO frequency band as the total mass decreases, which, in turn,
increases the discriminatory resolution of the matched-filter along the $q$ axis.
\begin{figure*}
\begin{center}
\includegraphics[width=\columnwidth]{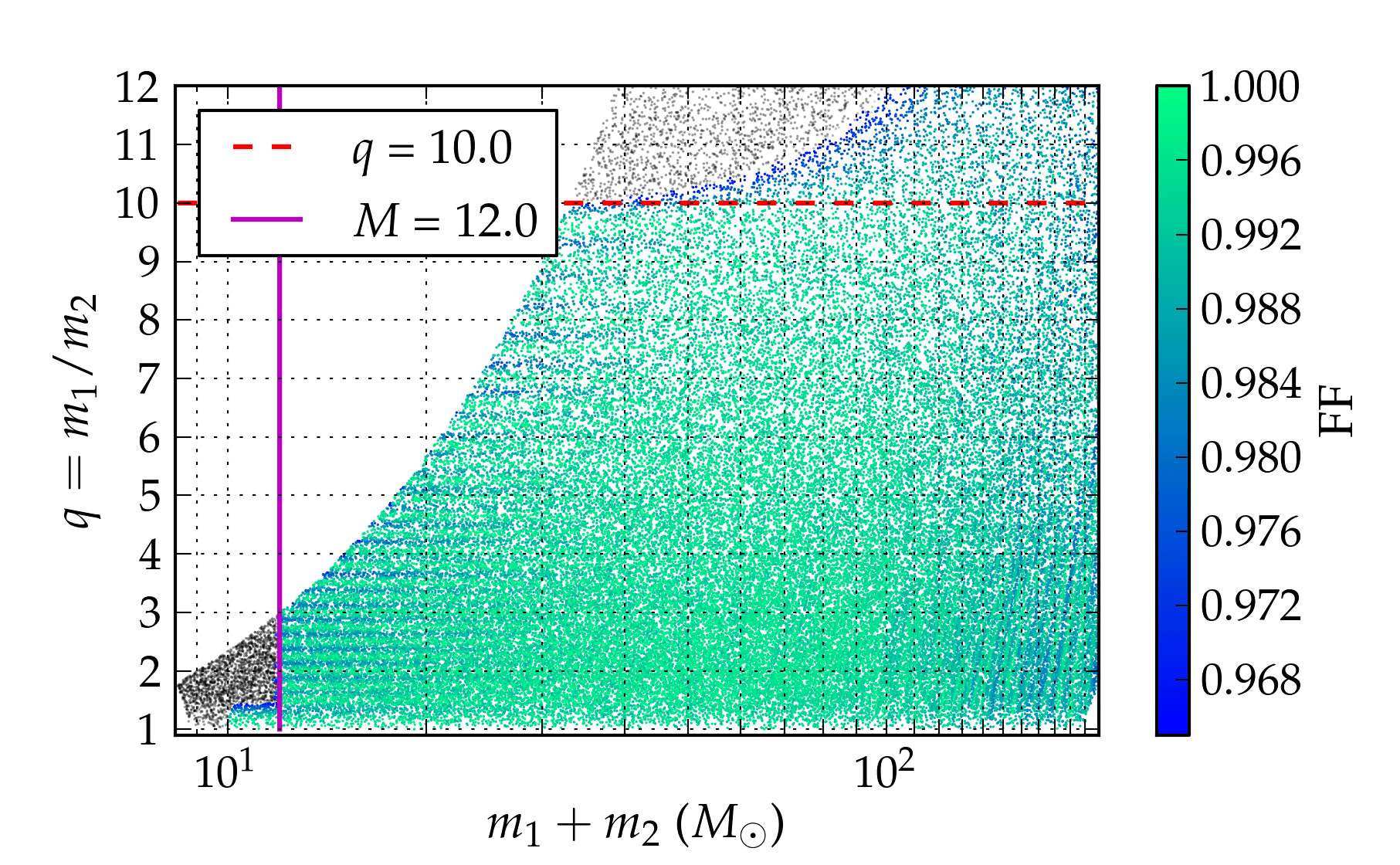}
\includegraphics[width=\columnwidth]{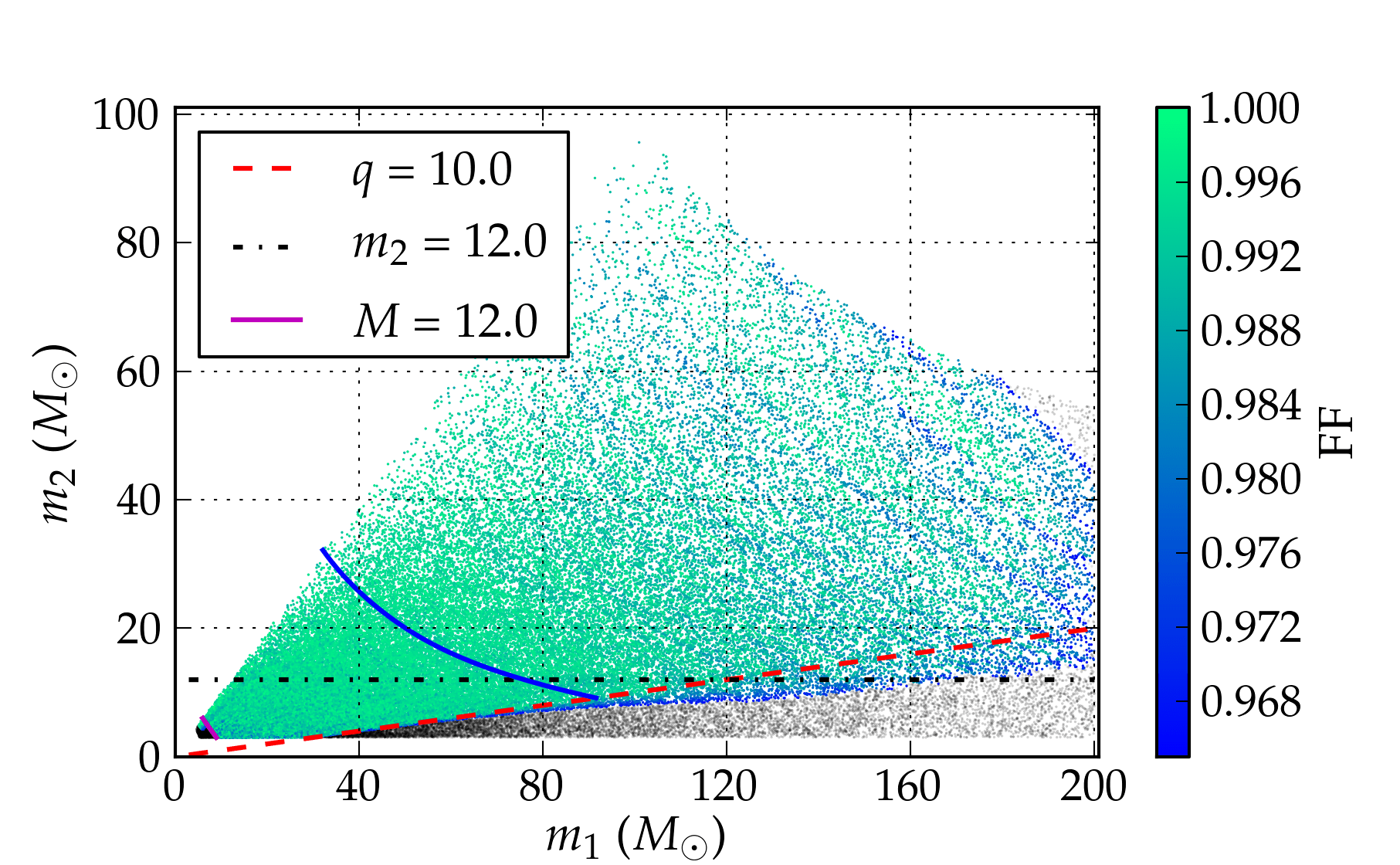}
\caption{\label{fig:templatebank_halfMassRatios}This figure shows
  fitting factors for a hybrid template bank which samples from the 26 mass
  ratios $q=1,1.5,1.75,2,..,9.6$, and allows coverage to masses down to 
  $m_1 + m_2 = 12M_{\odot}$ and $1\leq q\leq 10$, with a minimal-match of $98\%$
  at the lowest masses. 
  The left and right panel show the same on $M-q$ and $m_1-m_2$ axes, 
  respectively. The magenta lines, in both panels, shows the upper bound 
  in total mass, below which frequency-domain PN waveforms can be used to construct template banks for aLIGO
  searches~\cite{CompTemplates2009,Brown:2012nn}. The dash-dotted line
  in the right panel shows the lower mass limit on the smaller component object,
  to which a bank of currently available NR-PN hybrids can cover, i.e.
  $\mn(m_1,m_2)=12M_\odot$ (see Sec.~\ref{s1:NRpNhybridbank}). The blue (solid)
  curve in  the right panel gives the lower mass limit to which a bank of
  currently available NR waveforms can cover (see Sec.~\ref{s1:NRonlybank}).
  Thus, between the simulations listed in Table~\ref{table:fullqlist}, 
  and frequency domain PN waveforms, we can search for the entire range of 
  BBH masses.}
\end{center}
\end{figure*}
\begin{table}
\begin{tabular}{| c |}
\hline
$q\,(\equiv m_1/m_2)$ \\ \hline
1, 1.5, 1.75, \\
2, 2.25, 2.5, 2.75, \\
3, 3.25, 3.5, 3.8, \\
4.05, 4.35, 4.65, 4.95, \\
5.25, 5.55, 5.85, \\
6.2, 6.55, \\
7, 7.5, \\
8, 8.5, \\
9, 9.6 \\
\hline
\end{tabular}
\caption{List of mass-ratios, a template bank restricted to which will be effectual
over the region of the non-spinning BBH mass space where $m_1+m_2\gtrsim 12M_\odot$
and $1\leq q\leq 10$. The fraction of optimal SNR recovered by such a bank,
accounting for discreteness losses, remains above $98\%$. This is shown in
Fig.~\ref{fig:templatebank_halfMassRatios}.}
\label{table:fullqlist}
\end{table}
To determine the least set of mass-ratios which would sample the $q$ axis 
sufficiently densely at lower masses, we iteratively add mass-ratios to the 
allowed set and test banks restricted to sample from it. We find that, 
constrained to the set $\mathcal{S}_q$ given in Table~\ref{table:fullqlist},
a template bank can be constructed that has a minimal match of $98\%$
at the lowest masses. The left panel of Fig.~\ref{fig:templatebank_halfMassRatios}
shows the loss in SNR due to bank grid coarseness, i.e. $1-\Gamma_\mathrm{bank}$. 
This loss remains below $2\%$ for mass-ratios $1\leq q\leq 10$, even at 
$M=12M_\odot$. This 
leaves a margin of $1.5\%$ for the hybrid mismatches that would incur due to 
the hybridization of the NR merger waveforms with long PN inspirals. The right 
panel of Fig.~\ref{fig:templatebank_halfMassRatios} shows the same data in the 
$m_1$-$m_2$ plane. In this figure, 
the region covered by the NR-only bank is above the blue (solid) curve, while 
that covered by a bank of the currently available NR-PN hybrids is above the
line of $m_2 = 12M_\odot$ (with $m_2\leq m_1$). The region from 
Ref.~\cite{Brown:2012nn,CompTemplates2009} that can be covered by PN templates 
is in the bottom
left corner, bounded by the magenta (solid) line. Our bank restricted to the
set of $26$ mass-ratios, as above, provides additional coverage for binaries
with $M\geq 12M_\odot$, $m_2\leq 12M_\odot$ and $1\leq q\leq 10$.
Thus between purely-PN and NR/NR-PN hybrid templates, we can construct 
effectual searches for non-spinning BBHs with $q\leq 10$.

\begin{figure*}
\begin{center}
\includegraphics[width=0.33\textwidth, trim=17 20 75 75]{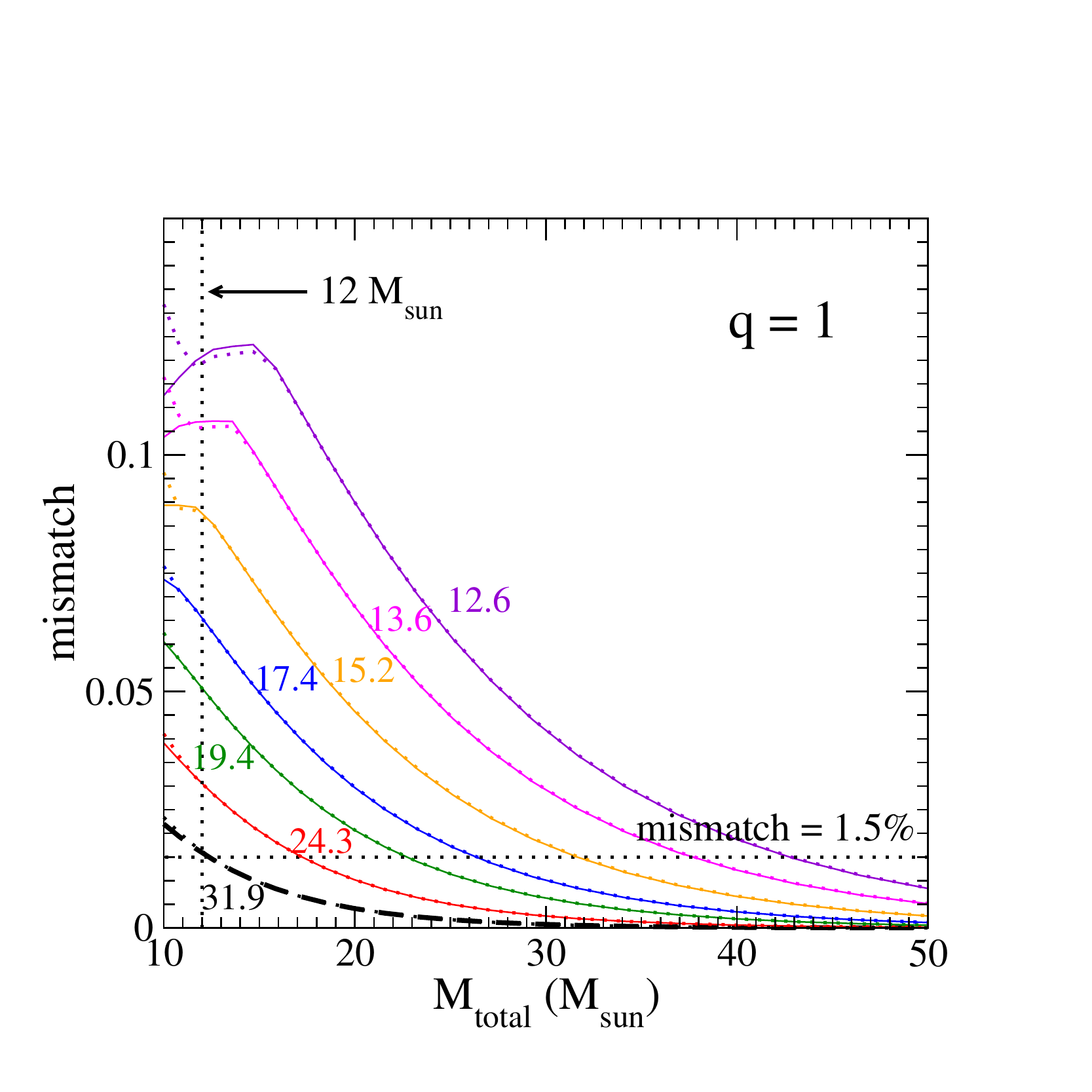}
\includegraphics[width=0.33\textwidth, trim=17 20 75 75]{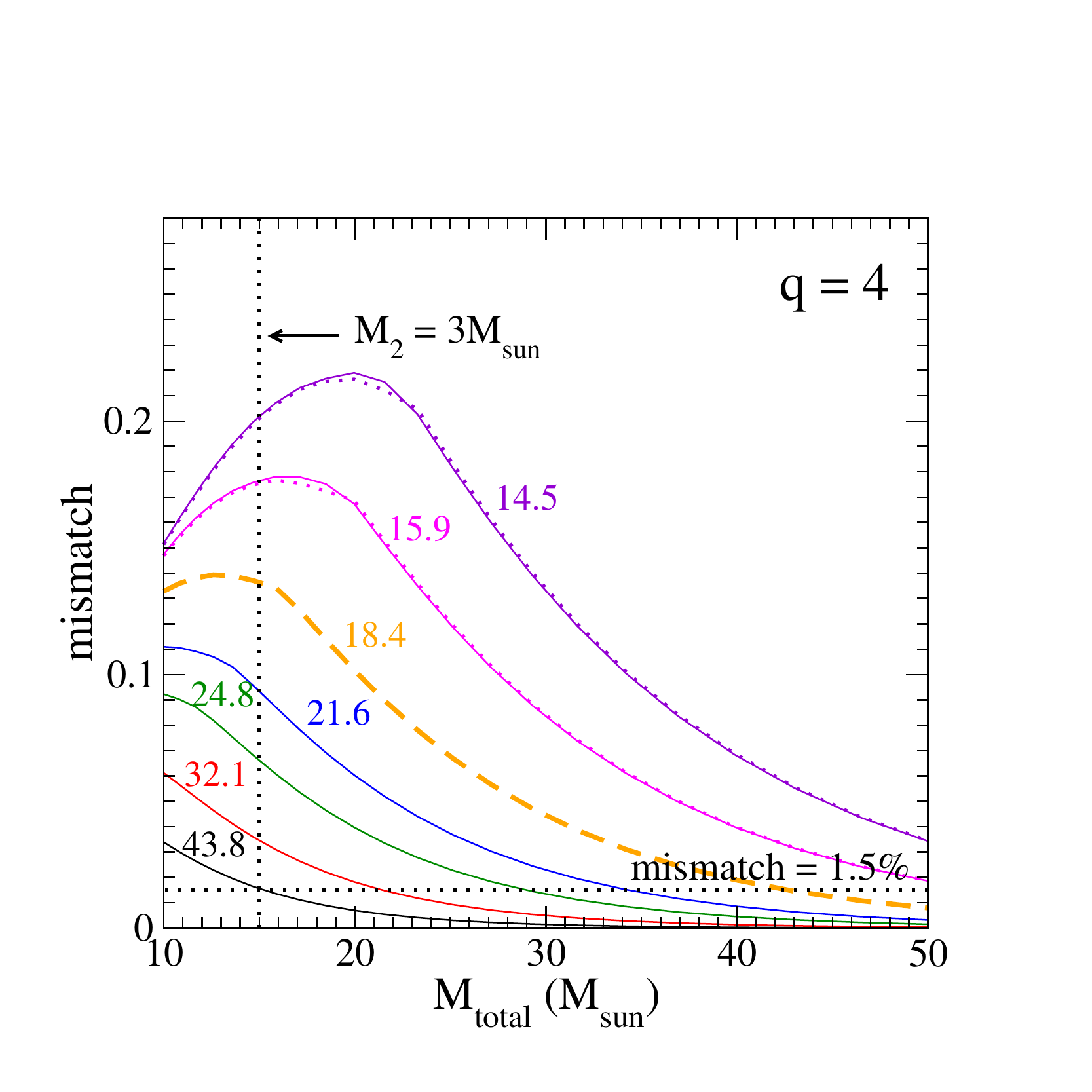}
\includegraphics[width=0.33\textwidth, trim=17 20 75 75]{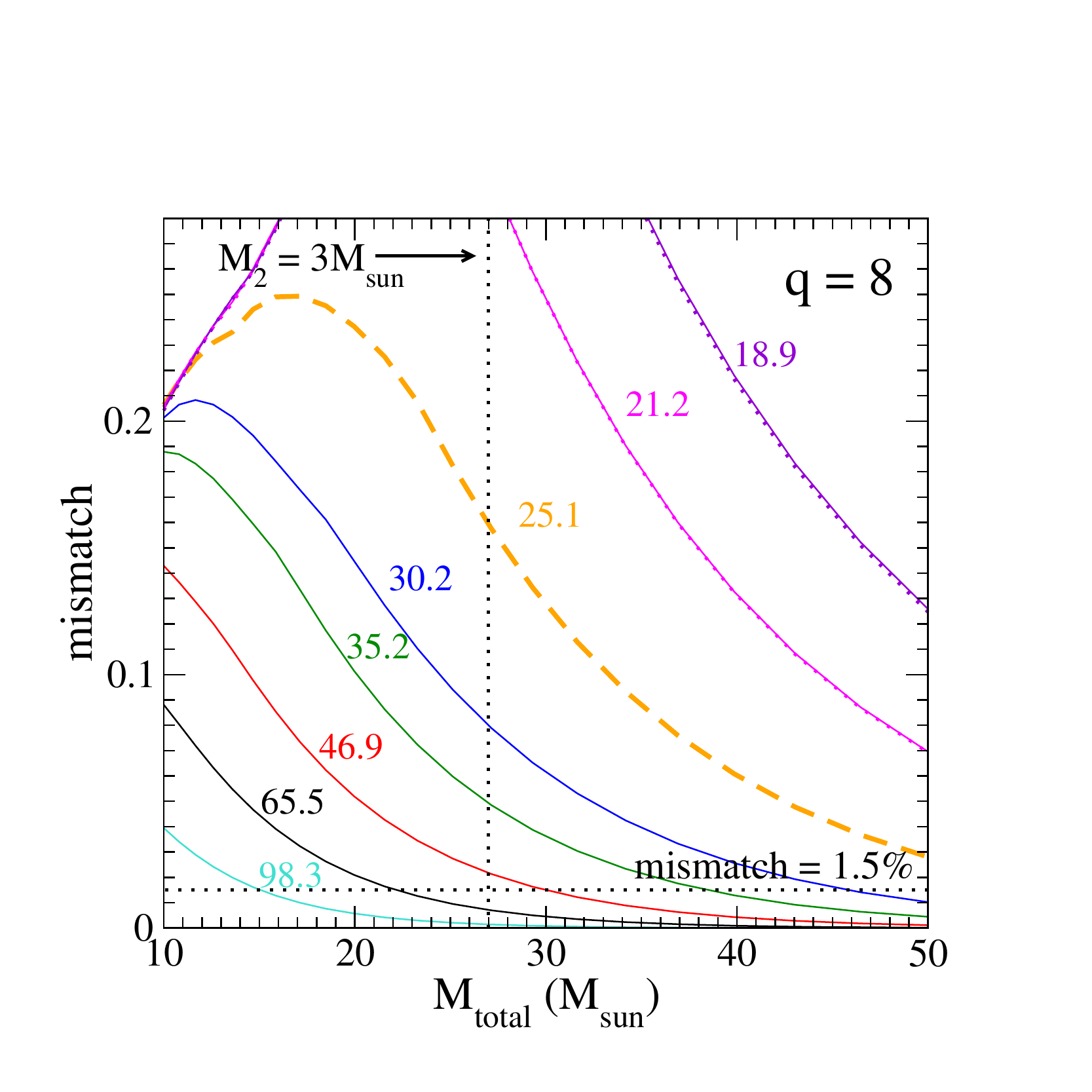}
\caption{\label{fig:maxmismatchVSmass} The maximum mismatch between
  different PN approximants for hybrid waveforms plotted against the
  total mass for at different matching frequencies ($M\omega_m$). The
  dotted lines indicate a mismatch of 1.5\% and a lower total mass
  limit, 12$M_\odot$ for $q=1$, and $M_2 = 3M_\odot$ for $q =
  4,8$. The thick dashed lines indicate the currently possible  
  matching frequency for hybrids based on the length of NR
  waveforms. The numbers next to each line indicate the number of
  orbits before merger where the PN and NR (or EOB) waveforms were 
  stitched together.} 
\end{center}
\end{figure*}

\begin{figure*}
\begin{center}
\includegraphics[width=\columnwidth, trim=17 25 75 25]{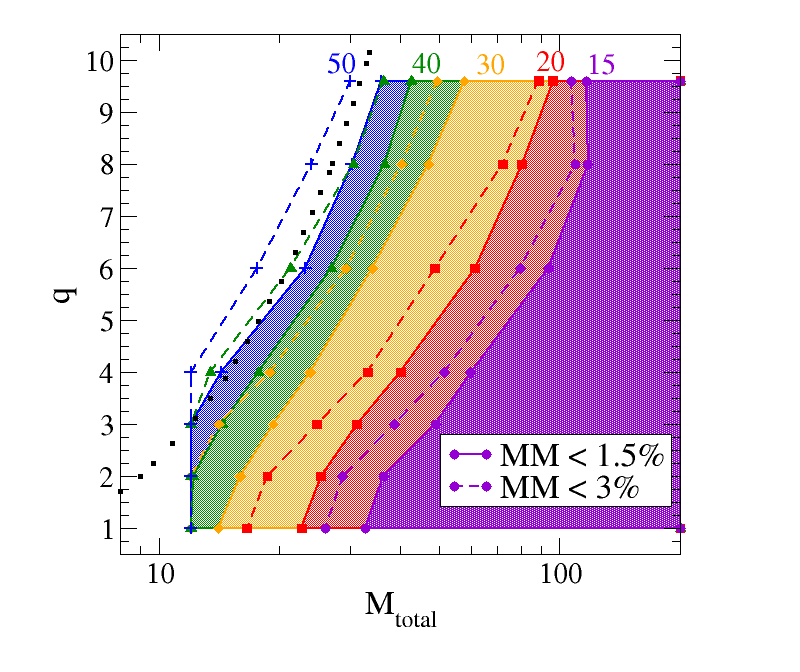}
\includegraphics[width=\columnwidth, trim=17 25 75 25]{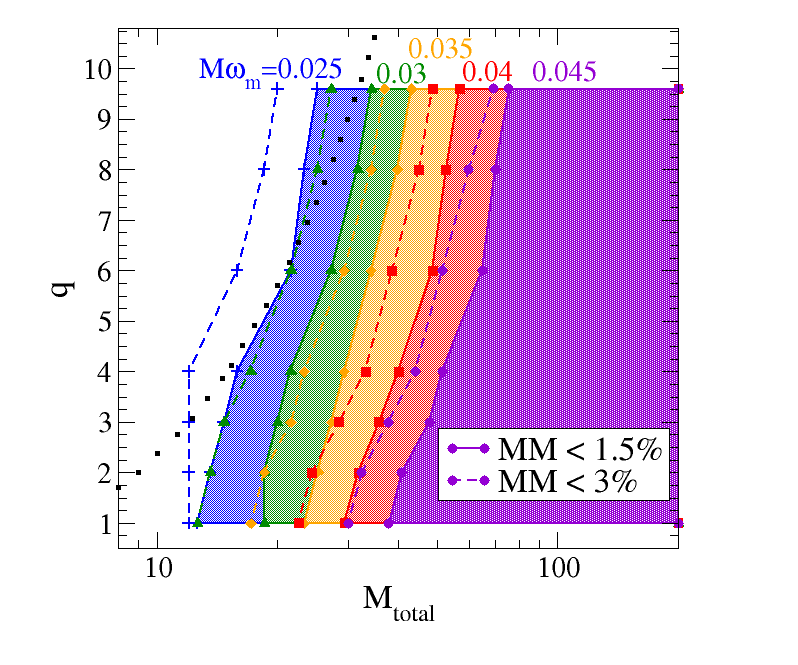}
\caption{\label{fig:NRorbits2merger} This plot shows the lower mass limit
  of a template bank constructed with hybrid waveforms in terms of the
number of NR orbits (left panel) and initial gravitational wave
frequency (right panel) needed to have a PN error below $1.5\%$ (solid
curves) or $3\%$ (dashed curves). The dotted line indicates the
lower total mass limit when one component mass is $3M_\odot$.}
\end{center}
\end{figure*}

Having the set of required mass-ratios ${\cal S}_q$ determined, 
we need to decide on the length requirements for the NR simulations,
in order to control the PN hybridization error.  
For a series of matching frequencies, we construct
NR-PN hybrids with Taylor\{T1,T2,T3,T4\} inspirals, and compute their
pairwise mismatch as a function of total mass. The maximum of
these mismatches serves a conservative bound on the PN-hybridization 
error for that hybrid (c.f. Eq.~(\ref{eq:GammaHybfinal})).
Fig.~\ref{fig:maxmismatchVSmass} shows the results of this calculation.
Each panel of
Fig.~\ref{fig:maxmismatchVSmass} focuses on one mass-ratio. Within
each panel, each line represents one matching-frequency, with lines
moving down toward earlier hybridization with smaller mismatches.
Because the hybridization frequency is not particularly intuitive, the
lines are labeled by the number of orbits of the NR portion of the
hybrid-waveform. For a short number of orbits this calculation is
indeed done with NR waveforms, whereas for large number of orbits, we
substitute EOBNRv2 waveforms. The dashed lines represent the earliest
one can match a NR+PN hybrid given the currently available NR
waveforms, and are the same as the $q = 1,4, \text{ and } 8$ lines in
Fig~\ref{fig:Current-NR-PN-Errors}. The solid curves show the results
using EOB hybrids, while the dotted curves (just barely visible) show
the results with NR hybrids. They are virtually identical, which is a
confirmation that EOB hybrids can act as a good proxy for NR
hybrids in this case. The horizontal dotted line
indicates a mismatch of 1.5\%, while the vertical dotted line shows a
lower mass limit for each mass ratio: $12M_\odot$ for $q=1$, which is
the point at which one can construct a template bank with only PN
inspirals, $15M_\odot$ for $q = 4$, and $27M_\odot$ for $q =8$, which
are the lower mass limits if both component masses are $\geq 3M_\odot$. 

Fig.~\ref{fig:NRorbits2merger} presents the information obtained in
the previous paragraph in a different way.  Given NR-PN hybrids with
$N$ orbits of NR, the shaded areas in the left panel of Fig.~\ref{fig:NRorbits2merger}
indicate the region of parameter space for which such hybrids have
hybridization errors smaller than $1.5\%$.  As before, we see that for
high masses, comparatively few NR orbits are sufficient (e.g. the
purple $N=15$ region), whereas lower total masses require increasingly
more NR orbits. The dashed lines indicate the region of parameter
space with hybrid error below $3\%$. The black dotted line designates
the point where one component mass is greater than $3M_\odot$, which
is a reasonable lower mass limit for a physical black hole. The right
panel shows this same analysis instead with initial GW frequency
indicated by the solid and dashed lines. Thus, for the region of
parameter space we're interested in, no more than $\sim 
50$ NR orbits, or an initial GW frequency of $M\omega = 0.025$ would
be necessary to construct a detection bank with  hybrid mismatches
below $1.5\%$.

%% file: conclusions.tex
The upgrades currently being installed to increase the sensitivity of the
ground based interferometric gravitational-wave detectors LIGO and
Virgo~\cite{Harry:2010zz,aVIRGO} are scheduled to complete within the next
two years. The second generation detectors will have a factor of $10$ better 
sensitivity across the sensitive frequency band, with the lower frequency
limit being pushed from $40$Hz down to $\sim 10$Hz. 
They will be able to detect GWs from stellar-mass BBHs up to distances of a few
Gpc, with the expected frequency of detection between 
$0.4 - 1000\, yr^{-1}$~\cite{LSCCBCRates2010}.

Gravitational-wave detection searches for BBHs operate by matched-filtering
the detector data against a bank of modeled waveform 
templates~\cite{Babak:2012zx,Sathyaprakash:1991mt, SathyaMetric2PN,
OwenTemplateSpacing,BabaketalBankPlacement,SathyaBankPlacementTauN,
Cokelaer:2007kx}.
Early LIGO-Virgo searches employed PN waveform template banks
that spanned only the inspiral phase of the
coalescence~\cite{Colaboration:2011nz,Abadie:2010yb,Abbott:2009qj,
Abbott:2009tt,Messaritaki:2005wv}.
Recent work has shown that a similar bank of PN templates would be
effectual for the advanced detectors, to detect non-spinning BBHs with 
$m_1 + m_2 \lesssim 12M_\odot$~\cite{Brown:2012nn,CompTemplates2009}.
Searches from the observation period between $2005-07$ and $2009-10$ 
employed templates that also included the late-inspiral, merger and 
ringdown phases of binary coalescence~\cite{Abadie:2011kd,Aasi:2012rja}. 

Recent advancements in Numerical Relativity have led to high-accuracy 
simulations of the late-inspiral and mergers of BBHs. The multi-domain SpEC
code~\cite{spec} has been used to perform simulations for non-spinning 
binaries with mass-ratios
$q=1,2,3,4,6,8$~\cite{Buchman:2012dw,Mroue:2012kv,Mroue:2013inPrep}. 
Owing to their high computational complexity, the length of these
simulations varies between $15-33$ orbits. Accurate modeling of the
late-inspiral and merger phases is important for stellar mass BBHs,
as they merge at frequencies that the advanced detectors would be 
sensitive to~\cite{Brown:2012nn}. Analytic models, like those within the 
Effective-One-Body formalism, have been calibrated to the NR simulations 
to increase their accuracy during these 
phases~\cite{EOBOriginalBuonannoDamour,EOBNRdevel01,BuonannoEOBv2Main,
Taracchini:2012ig}. Other independent models have also been developed 
using information from NR simulations and their hybrids~\cite{NRAR:home,
Ajith:2007qp,Santamaria:2010yb,Huerta:2012zy}. An alternate derived 
prescription is that of NR+PN hybrid waveforms, that
are constructed by joining long PN early-inspirals and late-inspiral-merger
simulations from NR~\cite{Boyle:2011dy,MacDonald:2011ne,MacDonald:2012mp,
Ohme:2011zm,Hannam:2010ky}.

NR has long sought to contribute template banks for gravitational-wave
searches. Due to the restrictions on the length and number of NR waveforms,
this has been conventionally pursued by calibrating intermediary
waveform models, and
using those for search templates. In this paper, we explore the alternative
of using NR waveforms and their hybrids directly in template banks.
We demonstrate the feasibility of this idea for non-spinning binaries,
and extending it to spinning binaries would be the subject of a future
work. We find that with only six non-spinning NR simulations, we can 
cover down to $m_{1,2}\gtrsim 12M_\odot$. We show that with
$26$ additional NR simulations, we can complete the non-spinning template
banks down to $M\simeq 12M_\odot$, below which existing PN waveforms 
have been shown to suffice for aLIGO. From template bank accuracy 
requirements, we are able to put a bound on the required length and 
initial GW frequencies for the new simulations. This method can therefore
be used to lay down the parameters for future simulations. 

First, we construct a bank for using pure-NR waveforms as templates, 
using a stochastic algorithm similar to Ref.~\cite{Harry:2009ea,
Ajith:2012mn,Manca:2009xw}. The filter templates are constrained to 
mass-ratios for which we have NR simulations available, i.e. 
$q=1,2,3,4,6,8$. 
We assume that the simulations available to us are $\geq 20$ orbits in
length. To test the bank, we simulate a population of $100,000$ BBH 
signals and filter them through the bank. The signals and templates 
are both modeled with the EOBNRv2 model~\cite{BuonannoEOBv2Main}. We 
demonstrate that this bank is indeed effectual and recovers 
$\geq 97\%$ of the optimal SNR for GWs from BBHs with mass-ratios 
$1\leq q\leq 10$ and chirp-mass 
$\mathcal{M}_c\equiv (m_1+m_2)^{-1/5}(m_1 m_2)^{3/5}$ above $27M_\odot$. 
Fig.~\ref{fig:bank001_01_match} shows this fraction at different 
simulated points over the binary mass space. With an additional 
simulation for $q=9.2$, we are able to extend the coverage to higher
mass-ratios. We show that a bank viable for NR waveform templates for
$q=1,2,3,4,6,9.2$, would recover $\geq 97\%$ of the optimal SNR
for BBHs with $10\leq q\leq 11$. The SNR recovery fraction from
such a bank is shown in Fig.~\ref{fig:bank006_01_match}.

Second, we construct effectual banks for currently available
NR-PN hybrid waveform templates. We derive a bound on waveform model
errors, which is independent of analytical models and can be used
to independently assess the errors of such models (see
Sec.~\ref{s1:quantifyingerrors} for details). This allows us to estimate
the hybrid waveform mismatches due to PN error, which are negligible
at high masses, and become significant at lower binary masses. We
take their contribution to the SNR loss into account while 
characterizing template banks. For hybrid banks, we demonstrate and 
compare two independent algorithms of template bank construction. 
First, we stochastically
place a bank grid, as for the purely-NR template bank. Second, we lay down
independent sub-banks for each mass-ratio, with a fixed overlap between
neighboring templates, and take their union as the final bank. 
To test these banks, we simulate a population of $100,000$ BBH signals
and filter them through each. We simulate the GW signals and the 
templates using the recently developed EOBNRv2 model~\cite{BuonannoEOBv2Main}. 
The fraction of the optimal SNR recovered by the two banks, before and after
accounting for the hybrid errors, are shown in the left and right panels 
of Fig.~\ref{fig:Current-hybrids-stochastic-FF} and 
Fig.~\ref{fig:Current-hybrids-FF} (respectively).
We observe that for BBHs with $m_{1,2}\geq 12M_\odot$ hybrid template 
banks will recover $\geq 96.5\%$ of the optimal SNR.  For testing the
robustness of our conclusions, we also test the banks using TaylorT4+NR
hybrid templates. The SNR recovery from a bank of these is shown in
Fig.~\ref{fig:Current-real-hybrids-FF}. We conclude that, 
the currently available NR+PN hybrid waveforms can be used as templates in 
a matched-filtering search for GWs from BBHs with $m_{1,2}\geq 12M_\odot$
and $1\leq q\leq 10$. The number of templates required to provide coverage
over this region was found to be comparable to a bank constructed using 
the second-order post-Newtonian TaylorF2 hexagonal template placement 
method~\cite{SathyaBankPlacementTauN,BabaketalBankPlacement,
SathyaMetric2PN,Cokelaer:2007kx}.
The two algorithms we demonstrate yield grids of $667$ and $627$ templates,
respectively; while the metric based placement method yields a grid of $522$ 
and $736$ templates, for $97\%$ and $98\%$ minimal match, respectively.

At lower mass, the length of the waveform in the sensitive frequency
band of the detectors increases, increasing the resolution of 
the matched-filter. We therefore see regions of undercoverage 
between mass ratios for which we have NR/hybrid templates 
(see, e.g. Fig.~\ref{fig:Current-hybrids-FF} at the left edge).
For $M\lesssim 12M_\odot$, existing PN waveforms were shown to 
be sufficient for aLIGO searches. 
We find the additional simulations that would be needed to extend
the hybrid tempalte bank down to $12M_\odot$. We show that a bank
of hybrids restricted to the $26$ mass-ratios listed in 
Table~\ref{table:fullqlist} would be sufficiently dense at $12M_\odot$.
This demonstrates that the method proposed here can be used 
to decide which NR simulations should be prioritized for the purpose
of the GW detection problem.
By filtering a population of $100,000$ BBH signals
through this bank, we show that the SNR loss due to its discreteness
stays below $2\%$ over the entire relevant range of masses.
The fraction of optimal SNR recovered is shown in 
Fig.~\ref{fig:templatebank_halfMassRatios}. Constraining the 
detection rate loss below $10\%$ requires that detection template 
banks recover more than $96.5\%$ of the optimal SNR. Therefore
our bank would need hybrids with hybridization mismatches below 
$1.5\%$. From this accuracy requirement, we obtain the length
requirement for all the $26$ simulations. This is depicted in the 
left panel of Fig.~\ref{fig:NRorbits2merger}, where we show the region
of the mass space that can be covered with hybrids, as the length of 
their NR portion varies. We find that for $1\leq q\leq 10$  the 
new simulations should be about $50$ orbits in length. In the right 
panel of Fig.~\ref{fig:NRorbits2merger} we show the corresponding
initial GW frequencies. The requirement of $\sim 50$ orbit long NR
simulations is ambitious, but certainly feasible with the current BBH 
simulation technology~\cite{BelaLongSimulation}.

In summary, we refer to the right panel of
Fig.~\ref{fig:templatebank_halfMassRatios}.
The region above the dashed (red) line and above the solid (blue) line can
be covered with a bank of purely-NR waveforms currently available. The region
above the dashed (red) and the dash-dotted (black) line can be covered with
the same simulations hybridized to long PN inspirals. With an additional set
of NR simulations summarized in Table~\ref{table:fullqlist}, the coverage
of the bank can be extended down to the magenta (solid) line in the lower
left corner of the figure. Thus between hybrids and PN waveforms, 
we can cover the entire non-spinning BBH space. The ability to use hybrid
waveforms within the software infrastructure of the LIGO-Virgo collaboration 
has been demonstrated in the NINJA-2 collaboration~\cite{NINJA2:2013inPrep}.
The template banks we present here can be directly used in aLIGO searches. 
This work will be most useful when extended to aligned spin and 
precessing binaries~\cite{Boyle:2013nka,Schmidt:2012rh}, which is the
subject of a future work.

The detector noise power is modeled using the zero-detuning high-power 
noise curve for Advanced LIGO~\cite{aLIGONoiseCurve}.
The construction of our template banks is sensitive to the breadth
of the frequency range that the detector would be 
sensitive to. The noise curve we use is the broad-band final design 
sensitivity estimate. For lower sensitivities
at the low/high frequencies, our results would become more conservative, 
i.e. the template banks would over-cover (and not under-cover).

We finally note that in this paper we have only considered the
dominant $(2,2)$ mode of the spherical decomposition of the 
gravitational waveform. For high mass-ratios and high binary masses,
other modes would also become important, both for spinning as well
as non-spinning black hole binaries~\cite{Pekowsky:2012sr,
Brown:2012nn,Capano:2013inPrep}.
Thus, in future work, it would be relevant to examine the
sub-dominant modes of the gravitational waves. Lastly, though we have
looked at the feasibility of using this template bank for Advanced
LIGO as a single detector, this instrument will be part of a network
of detectors, which comes with increased sensitivity and sky
localization. For this reason, in subsequent studies it would be
useful to consider a network of detectors.